\documentclass[letterpaper,journal]{IEEEtran}      

\usepackage{amsmath} 
\usepackage{amssymb}  
\usepackage{cite}
\usepackage{color}
\newcommand{\abs}[1]{\lvert#1\rvert}

\usepackage[dvips]{graphicx}

\newtheorem{theorem}{Theorem}[section]
\newtheorem{proposition}[theorem]{Proposition}
\newtheorem{lemma}[theorem]{Lemma}
\newtheorem{remark}[theorem]{Remark}
\newtheorem{corollary}[theorem]{Corollary}
\newtheorem{definition}[theorem]{Definition}
\newtheorem{algorithm}[theorem]{Algorithm}

\usepackage{enumerate}
\usepackage{subfigure}

\newcommand\floor[1]{\lfloor#1\rfloor}
\newcommand\ceil[1]{\lceil#1\rceil}

\begin{document}

\title{\LARGE \bf 
Resilient Randomized Quantized Consensus}

\author{Seyed Mehran Dibaji, Hideaki Ishii, and Roberto Tempo%
\thanks{S.~M.~Dibaji is with the Department of Mechanical Engineering, 
Massachusetts Institute of Technology, 
Cambridge, MA 02139, USA.
E-mail: {dibaji@mit.edu}}
\thanks{H.~Ishii is with the Department of Computer Science, 
Tokyo Institute of Technology, Yokohama,
226-8502, Japan.
E-mail: {ishii@c.titech.ac.jp}}
\thanks{R. Tempo is with CNR-IEIIT, Politecnico di Torino, 
10129 Torino, Italy.
E-mail: {roberto.tempo@polito.it}}
\thanks{This work was supported in part by the JST CREST Grant No. JPMJCR15K3,
by the CNR-JST International Joint Lab COOPS, 
and by JSPS under Grant-in-Aid for Scientific
Research Grant No.~15H04020.}}

\maketitle
\thispagestyle{empty}
\pagestyle{empty}

\begin{abstract}
We consider the problem of multi-agent consensus where some agents 
are subject to faults/attacks and might make updates arbitrarily. 
The network consists of agents taking integer-valued (i.e., quantized) 
states under directed communication links. 
The goal of the healthy normal agents is to form consensus 
in their state values, which may be disturbed by the non-normal, 
malicious agents. 
We develop update schemes to be equipped by the normal agents 
whose interactions are asynchronous and subject to non-uniform 
and time-varying time delays. 
In particular, we employ a variant of the so-called mean 
subsequence reduced (MSR) algorithms, which have been long 
studied in computer science, 
where each normal agent ignores extreme values from its neighbors. 
We solve the resilient quantized consensus problems in the 
presence of totally/locally bounded adversarial agents
and provide necessary and sufficient conditions
in terms of the connectivity notion of graph robustness. 
Furthermore, it will be shown that randomization is 
essential both in quantization and in the updating times 
when normal agents interact in an asynchronous manner. 
The results are examined through a numerical example.
\end{abstract}


\section{Introduction}\label{Sect: Introduction}

In recent years, studies on networked control systems with an emphasis 
on cyber security have received a growing attention. Due to 
the use of general purpose networks in large-scale control systems, 
security against malicious intrusions is becoming a key issue. 
One of the fundamental problems is the so-called resilient consensus,
which is the multi-agent consensus problem (e.g., \cite{mesbahi,RenCao:11}) 
in the presence of agents subject to faults and attacks. 
In such problems, non-faulty agents collaborate with each other 
to attain global agreement while the faulty agents may make updates
arbitrarily, which can affect the behavior of the normal agents. 
\textcolor{black}{%
The resilient versions of the consensus algorithms provide the means 
to protect multi-agent systems from faults and cyber attacks 
in applications such as autonomous mobile agents and sensor networks.} 

More specifically, in resilient consensus problems, each non-faulty (or \textit{normal})
agent is assumed to be aware of only local information available from its
neighbors regarding their states. 
In contrast, malicious agents may have more global information
regarding the behavior of normal agents, by even exchanging 
information over links not present in the network.
The objective in these problems is to design distributed control protocols 
for the normal agents to achieve consensus among themselves
and to establish conditions under which such resilient consensus can
be attained. 

Resilient consensus has a rich history in the area of distributed algorithms 
in computer science (see, e.g., \cite{Lynch} and the references therein).
It is however highlighted 
that many works there deal with networks of complete graphs.
\textcolor{black}{%
This may be due to the fact that agents in this area are often models
of computer terminals connected over wired networks, performing 
load balancing through consensus algorithms.}
From the current perspective of multi-agent consensus in systems and control, 
it is of interest to find the minimum requirement on networks to 
warrant resiliency against faults and attacks. 
\textcolor{black}{%
In this paper, we will address this fundamental issue in a problem 
setting that is in accordance with the computer science literature.}

In particular, we consider resilient consensus problems in the setting
where the agents take integer-valued states and the underlying network
for their communication is directed and non-complete. 
Quantized consensus has been motivated by concerns on limited capabilities 
in communications and computations of the agents, and various studies have
recently been carried out for the case without any malicious agents 
\cite{Aysal,
CaiIshii,
CarliFagnaniFrascaZampieri,
ChamieLiuBasar,
FrascaZampieri,
GravelleMartinez,
KarMoura,
KashyapBasarSrikant,
LavaeiMurray,
Li,
NedOlshOzdTsitsilis}. 
For enhancing resiliency,  
we employ a distributed update scheme in which each normal 
agent ignores its neighbors whose states appear extreme and unsafe 
in the sense that they differ the most from its own.
By assuming that the maximum number $f$ of malicious agents in the network
is known, each normal agent neglects up to $f$ largest and up 
to $f$ smallest values from its neighbors. 
Such update schemes are often called \textit{mean subsequence reduced} (MSR) type 
algorithms. 
We study the problem for both synchronous and asynchronous updates
for the normal agents and obtain necessary and sufficient conditions in
the underlying network structure for the agents' communication. 

For resilient consensus in our problem setting, the critical
notion for network structures is 
called \textit{graph robustness}. It is a measure of connectivity within
a graph and characterizes how well groups within the network are 
connected via multiple paths.
It was first introduced by \cite{LeBlanc} for resilient consensus 
in the real-valued states case 
with first-order agent dynamics and then was further explored in
\cite{DibIshiiIFAC,Tseng}. 
\textcolor{black}{%
The resilient approach has moreover been extended to networks with
higher-order agents. 
In \cite{DibIshiiSCL,DibIsh:aut16}, we derived similar results for 
vehicle agents having second-order dynamics. 
Furthermore, an MSR-type algorithm is constructed for the application of 
clock synchronization in wireless sensor networks \cite{KikDibIsh:16},
where each node is equipped with an update scheme of two states.}

Our viewpoint has been
motivated by the recent literature in control on multi-agent systems and
has led us to introduce features in the update schemes and communication 
delays different from those in computer science. It is evident 
that in control, real-time aspects in algorithms have more significance. 
On the other hand, the condition on graph structures may be relaxed 
by following an approach based on fault detection and isolation 
techniques; this will however require to equip
all agents with banks of observers (e.g., \cite{pasqdorfler}), 
which may be difficult to implement considering the limited 
computational resources on the agents.

We emphasize that the proposed update schemes employ probabilistic 
techniques, which turn out to be very important 
in the implementation of our algorithms. 
For general references on randomization-based algorithms
in systems and control, we refer to \cite{TempoBook,TempoIshii}.
Randomization is introduced for two objectives.
One is in quantization. Since the update schemes involve 
iterative weighted averaging of integer-valued states, the resulting real number must be rounded by a quantizer. We employ quantizers that perform randomization. They have an 
effect similar to dithering, commonly used in speech and video processing,
for controlling statistical properties introduced by quantization noises 
\cite{WanLipVan:00}.
It will be shown that deterministic quantizers are not sufficient. 
Related phenomena have been found in
\cite{Aysal,CarliFagnaniFrascaZampieri,ChamieLiuBasar,KarMoura} for
quantized average consensus problems over undirected graphs with no 
malicious agent.

\textcolor{black}{%
The other part where randomization is utilized is 
in the updating times of the agents. This is sometimes called gossiping 
(see, e.g., \cite{KashyapBasarSrikant,RavazziFrascaTempoIshii,TempoIshii}),
but has not been exploited in the recent studies of resilient consensus
discussed above.
Our novel finding is 
that for asynchronous updates in the normal agents,
randomized updating is essential in establishing resilient consensus under the
minimal robustness requirement in the agent network.
The intuitive reason is that due to gossipping, the malicious agents have
less information about the normal agents and thus
cannot coordinate their behaviors well to misguide them.}
In particular, with randomization in updating times, the necessary and sufficient 
condition coincides with that for the synchronous counterpart.
We will show some scenarios under deterministic updates,
where consensus fails on graphs possessing sufficient connectivity for 
the synchronous updates case.

Our results suggest that randomization is critical in resilient consensus. 
We would like to mention an interesting connection to computer 
science; for more details, see \cite{Dibaji}. 
There, randomization is used as an important component for many algorithms 
to enhance efficiency \cite{MotwaniRaghavan}. 
This has in fact been well explored in the context of binary-valued
consensus problems when adversaries are present \cite{Lynch}.
For example, under asynchronous deterministic updates, even one agent 
halting its operation can make it impossible for the system to reach consensus \cite{FisLynPat:85}; 
however, the use of probabilistic techniques can relax the situation \cite{BenOr:83}. 
Another example is the so-called Byzantine general problem, 
where randomization is necessary to obtain algorithms scalable in 
convergence times with respect to the size of the networks \cite{MotwaniRaghavan,Rabin:83}. 
We note that our results can be applied to the binary state case 
by simply restricting the initial values to 0 and 1.

Regarding the interaction among agents, 
we consider two cases where the communications are immediate and also those
which experience non-uniform and time-varying delays.  
In both cases, we characterize the network structures in terms of robustness
and provide sufficient conditions and necessary conditions, which sometimes coincide. 
It however becomes clear that time delays bring further vulnerabilities into the
agent system. As a result, we find that additional connectivity is necessary
to achieve resilient consensus for both deterministic and probabilistic updating schemes. 
We show through a numerical example how a malicious
agent can exploit the delays in such a way that the normal agents 
become divided into groups. However, without delays in communication, 
this type of vulnerability can be prevented by means of randomization 
as discussed above.

Furthermore, as a side result, we show that when no malicious agent is 
present in the network (i.e., with $f=0$), the necessary and sufficient conditions
in our main results coincide, establishing the topological condition of 
having a spanning tree in the network for quantized consensus. 
Though this condition is well known in the literature of consensus, 
in the particular problem setting of quantized states and directed
networks, 
the result is new to the best of our knowledge. 

The paper is organized as follows: 
In Section~\ref{Sect: Preliminaries}, we present preliminary material 
and introduce the problem setting. 
Section~\ref{Sect:synch} is devoted to the quantized 
consensus problem in the presence of malicious agents when
the update schemes for the normal agents are synchronous. 
Then, in Section~\ref{Sect:Asynch}, the asynchronous
counterpart without any delay is analyzed, where randomization 
in update times is proven to be useful. In Section~\ref{Sect: DELAY}, 
the problem is solved in the presence of delays in communication among the agents. 
Further, Section~\ref{Sect: Simulations} provides numerical examples to 
illustrate the effectiveness of the proposed algorithms. 
Finally, in Section~\ref{Sect: Conclusion}, we discuss concluding remarks and 
future directions. This paper is an expanded version of 
the conference papers
\cite{DibIshiiTempoACC16,DibIshiiTempoCDC16} and
contains the full proofs of the theoretical results 
with extended discussions on the role of randomizations,
the relations to the literature in computer science,
and the simulation results.

\section{Preliminaries and Problem Setup}\label{Sect: Preliminaries}

In this section, we first provide preliminary material on graphs and then
introduce the basic problem setting for the resilient consensus problems
studied in this paper.

\subsection{Graph Theory Notions}\label{graphnot}
In this section, we recall some concepts on graphs \cite{mesbahi}.

A weighted directed graph with $n$ agents is defined as a triple
$\mathcal{G}=\left(\mathcal{V},\mathcal{E},A\right )$ with the set of nodes $\mathcal{V}=\{1,\ldots,n\}$, the set of edges  $\mathcal{E}\subseteq \mathcal{V}\times\mathcal{V}$, and the adjacency matrix $A\in \mathbb{R}^{n\times n}$.
The edge $(j,i)\in \mathcal{E}$ means that node $i$ has access to the 
information of node $j$. We do not allow self-loops, that is, $(i,i)\notin \mathcal{E}$.
If each node has an edge to all other nodes, the corresponding graph is said to be complete.
For node $i$, the set of its neighbors consists of all nodes which have directed edges toward $i$, and it is denoted by $\mathcal{N}_i=\{j:(j,i)\in \mathcal{E}\}$.  
The degree of node $i$ is the number of its neighbors and is denoted by ${d}_i=\abs{\mathcal{N}_i}$.  
If the edge $(j,i)$ exists, the associated entry $a_{ij}$ in the adjacency matrix $A$ is in $(\alpha ,1)$, and otherwise $a_{ij}$ is zero, where $0<\alpha <1$. We assume that $\sum_{j=1,j\neq i}^{n} a_{ij} < 1$.
Let $L=[{l}_{ij}]$ be the Laplacian matrix of $\mathcal{G}$ 
whose entries are defined as $l_{ii}=\sum_{j=1,j\neq i}^{n}a_{ij}$ 
and $l_{ij}=-a_{ij}$ for $i\neq j$. 
It is clear that the sum of the elements of each row of the Laplacian matrix is zero. 

A path from node $i_1$ to $i_p$ is a sequence of distinct nodes $(i_1,i_2,\ldots,i_p)$, where $(i_m,i_{m+1}) \in \mathcal{E}$ for $m=1, \ldots,p-1$. If for all distinct nodes $i$ and $j$, there is a path from $i$ to $j$, the graph is called strongly connected. A directed graph is said to have a directed spanning tree if
there is a node having a path to every other node in the graph. 

For the algorithms proposed in this paper, we characterize 
topological properties of networks in terms of graph robustness. 
It measures the connectivity in a graph by showing how well groups within 
the network are connected over different paths.
It was first introduced by \cite{LeBlanc} for analysis of resilient consensus of real-valued first-order multi-agent systems. Related works include
\cite{DibIshiiIFAC} which studied the case with delays in communication 
and \cite{DibIshiiSCL,DibIsh:aut16} for the case of agents 
with real-valued second-order dynamics. 
We use the more general notion of $(r,s)$-robust graphs, which plays an important 
role to obtain a tight necessary and sufficient condition. 

\begin{definition}\label{Def: RobustGraph}\rm
The graph $\mathcal{G}= (\mathcal{V},\mathcal{E},A)$ is $(r,s)$-robust $(r,s<n)$ if for every pair of nonempty disjoint subsets $\mathcal{V}_1,\mathcal{V}_2 \subset \mathcal{V}$, at least one of the following conditions holds:
\begin{enumerate}
\item ${\mathcal{X}_{\mathcal{V}_1}^r} ={\mathcal{V}_1}$,
\item ${\mathcal{X}_{\mathcal{V}_2}^r}={\mathcal{V}_2}$,
\item $\abs{\mathcal{X}_{\mathcal{V}_1}^r} +\abs{\mathcal{X}_{\mathcal{V}_2}^r} \geq s$,
\end{enumerate}
where $\mathcal{X}^r_{\mathcal{V}_{\ell}}$ is the set of nodes 
in ${\mathcal{V}_{\ell}}$ having at least $r$ incoming edges 
from outside ${\mathcal{V}_{\ell}}$.
\textcolor{black}{%
As the special case with $s=1$, graphs which are $(r,1)$-robust are 
called $r$-robust.}
\end{definition}

The following lemma provides a better understanding 
of robust graphs \cite{LeBlancPhD}.
Here, $\ceil y$ denotes the ceiling function and gives 
the smallest integer value greater than or equal to $y$.


\begin{lemma}\label{lemma: robustgraphs}\rm
	For an $(r,s)$-robust graph $\mathcal{G}$, the following hold:
	\begin{enumerate}[(i)]
		\item $\mathcal{G}$ is $(r',s')$-robust, where $0\leq r'\leq r$ and $1 \leq s'\leq s$, and in particular, it is $r$-robust.
		\item $\mathcal{G}$ is $(r-1,s+1)$-robust.
		\item $\mathcal{G}$ has a directed spanning tree. Moreover a graph is 1-robust if and only if it has a directed spanning tree.
		\item ${r} \leq \lceil {n / 2} \rceil$. Further, 
if $r=\lceil {n/2} \rceil $, $\mathcal{G}$ is a complete graph. 
	\end{enumerate}
	Moreover, a graph $\mathcal{G}$ is $(r,s)$-robust 
	if it is $(r+s-1)$-robust.
\end{lemma}

\begin{figure}[t]
	\centering
	\vspace*{3mm}
	\includegraphics[width=35mm]{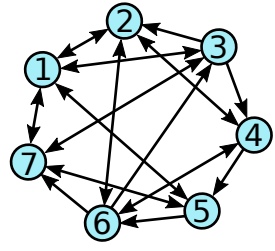}
	\vspace*{0mm}
	\caption{\textcolor{black}{A $(2,2)$-robust graph with seven nodes.}}
	\label{fig: 7node22robust}
\end{figure}

It is clear that $(r,s)$-robustness is stronger than $r$-robustness. 
The graph with seven nodes in Fig.~\ref{fig: 7node22robust} can be shown to be $(2,2)$-robust, but not $3$-robust.
In general, to determine whether a given graph
possesses a robustness property is computationally difficult because 
the problem involves combinatorial issues. 
It is known that certain random networks are robust when they are sufficiently connected \cite{MoradiPiraniSundaram,zhangfata,ZhaoYaganGligor}.

\subsection{Consensus among Integer-Valued Agents}

In the remaining of this section, we give the problem formulation 
of resilient consensus for the case without time delays
in the interactions among the agents. 

Consider a network of agents cooperating over the directed graph $\mathcal{G}=(\mathcal{V},\mathcal{E},A)$. Each agent applies a control rule consisting of its neighbors' state values to make updates by
\begin{equation}\label{eqn: GeneralDynamic}
x_i[k+1]=x_i[k]+u_i[k],
\end{equation}
where $x_i[k]$ and $u_i[k]$ represent the state and the control input
of agent $i$ at time $k$. 

To achieve consensus means that the agents converge to 
a globally common value. 
In typical consensus problems, the agents are real-valued 
(i.e., $x_i[k]\in \mathbb{R}$), and they approach consensus asymptotically. 
A well-known approach for updating is to apply 
the weighted average of the relative state 
values of the agent and its neighbors as
\begin{equation}\label{eqn: GeneralControl}
  u_i[k]=\sum_{j\in \mathcal{N}_i} a_{ij}[k] (x_j[k]-x_i[k]),
\end{equation}
where $a_{ij}[k]$ is the ${(i,j)}$th entry of the adjacency matrix $A[k]$ of the graph $\mathcal{G}[k]$ at time $k$. 

Here, we consider the situation where limited communication and memory
of the agents enforce them to take integer values.  
Hence, the states and the inputs are constrained as $x_i[k]\in\mathbb{Z}$ and
$u_i[k]\in\mathbb{Z}$
for $i\in\mathcal{V}$.

In the update rule, we employ the quantization function $Q:\mathbb{R} \to \mathbb{Z}$ 
to transform the real-valued input in \eqref{eqn: GeneralControl}
to an integer-valued one.
This is done in a probabilistic manner as \cite{Aysal} 
\begin{equation}
  Q(y) = \begin{cases}
              \floor y & \text{with probability $p(y)$},\\
              \ceil y  & \text{with probability $1-p(y)$},
           \end{cases}
\label{eqn:Q}
\end{equation}
where $p(y)= {\ceil y -y}$, and the floor
function \noindent$\floor y$ gives the greatest integer less than or equal 
to $y$.
Based on \eqref{eqn: GeneralControl}, 
the quantized control input of agent $i$ can be written as
\begin{equation}
 u_i[k] 
  = Q\biggl(
          \sum_{j\in \mathcal{N}_i} a_{ij}[k] (x_j[k]-x_i[k])
        \biggr).
\label{eqn:u_i}
\end{equation}
It is noteworthy that the probabilistic quantizer equipped 
on each agent is independent and determines whether to choose 
the floor or ceil function at each time. 
Thus, the control \eqref{eqn:u_i} can be implemented in
a distributed fashion.
Probabilistic quantizers have been studied in \cite{Aysal,CarliFagnaniFrascaZampieri,ChamieLiuBasar,KarMoura} for average
consensus of real-valued agent networks with quantized communications. 
Moreover, it can be shown that probabilistic quantization is equivalent to 
the well-known dithering \cite{Aysal}, which has a range of 
applications in digital signal processing. 

In our problem setting, we also introduce asynchrony in the updates. 
That is, at each time $k$, agent $i$ may or may not make an update
by applying its control $u_i[k]$. 
If an agent does not update its value, it keeps its previous value, 
i.e., $x_i[k+1]=x_i[k]$.
Denote by $\mathcal{U}[k]\subset\mathcal{V}$ 
the set of agents that update their values at time $k$.
The agent system is said to be \textit{synchronous} if 
$\mathcal{U}[k]=\mathcal{V}$ for all $k$, and otherwise 
it is \textit{asynchronous}. 
It is noted that the control in \eqref{eqn:u_i}
already represents the asynchronous case: 
If $i\notin\mathcal{U}[k]$, 
then $a_{ij}[k]=0$ for $j\in\mathcal{N}_i$.

 
The objective of the agent network is to achieve global consensus among 
all agents in a probabilistic sense
by applying the possibly asynchronous local control input \eqref{eqn:u_i}.
In this work, the network is assumed to have some misbehaving agents that 
do not follow the control in \eqref{eqn:u_i}. 
In the next subsection, we provide the required notions to study this case. 
Note that networks without any malicious agents form a special case;
for such normal networks, we obtain consensus conditions
as a direct consequence of the results obtained in this paper. 


\subsection{Resiliency Notions and Algorithm}\label{sec:alg1}


We introduce notions related to malicious agents and consensus in the presence of 
such agents \cite{DibIshiiIFAC,DibIshiiSCL,LeBlanc}.

\begin{definition}\label{Def: NormalNode}\rm
Agent $i$ is called normal if it updates its state based on the 
predefined control \eqref{eqn:u_i}.
Otherwise, it is called malicious
and may make arbitrary updates. The index set of malicious agents
is denoted by $\mathcal{M}\subset\mathcal{V}$. 
\end{definition}

The numbers of normal agents and malicious agents are denoted by $n_N$ and $n_M$, 
respectively. 
Also, the states for the normal agents and malicious agents
are given in vector forms as $x^N[k]\in\mathbb{Z}^{n_N}$ 
and $x^M[k]\in\mathbb{Z}^{n_M}$, respectively. 

\textcolor{black}{%
In the agent dynamics in \eqref{eqn: GeneralDynamic}, 
the control inputs for the normal agents are given by \eqref{eqn:u_i} 
while the malicious agents can choose their control arbitrarily, and hence
$u_i[k]$ for $i\in\mathcal{M}$ are left unspecified at this point. 
Thus, the update rules for the agents can be written 
in the following form:}
\begin{align}
 x_i[k+1] 
  &= \begin{cases}
      Q\biggl(
         \sum_{j\in \mathcal{N}_i\cup\{i\}} w_{ij}[k] x_j[k]
       \biggr) 
      & \text{if $i\in\mathcal{V}\setminus\mathcal{M}$},\\
      x_i[k] + u_i[k] & \text{if $i\in\mathcal{M}$}.
     \end{cases}
\label{eqn: ProbQuanGenUpdatesNONDELAY}
\end{align}
Here, we have $W[k]=(w_{ij}[k])=I-L[k]$, where
$L[k]$ is the Laplacian matrix associated with the graph $\mathcal{G}[k]$. 
From the definition of the Laplacian matrix, 
it follows that $W[k]$ is row stochastic, that is, 
all entries are nonnegative and each row sum is equal
to one; moreover, each positive entry of $W[k]$ is lower bounded by $\alpha$. 
Thus, each normal agent makes an update by a quantized convex combination 
of its neighbors' state values. 

With respect to the number of misbehaving agents in the network, 
we assume that an upper bound denoted by $f$ is known to all
agents. 
More precisely, we consider the two cases defined below. 

\begin{definition}\rm
The network is said to be $f$-total malicious if the number $n_M$ of faulty 
agents in the entire network is at most $f$, i.e., $n_M \leq f$.
On the other hand, the network is called $f$-local malicious if 
for each normal agent $i \in \mathcal{V}\setminus \mathcal{M}$, 
the number $n_M$ of faulty agents in its neighborhood is at most $f$, i.e., 
$\abs{\mathcal{M}\cap\mathcal{N}_i} \leq f$. 
\end{definition} 

We now introduce the notion of resilient consensus for the network of probabilistic quantized 
agents in the presence of misbehaving agents. 

\begin{definition}\label{Def: ResilientConsensus}\rm
If for any initial states, any malicious inputs, and 
any possible set of malicious agents, the following 
conditions are met, then the network is said to reach 
resilient quantized consensus:
\begin{enumerate}
\item Safety condition:
For each set of initial values of the normal agents, there exists a set $\mathcal{S}$ such that for all normal agents $i \in \mathcal{V}\setminus \mathcal{M}$, 
it holds that $x_i[k]\in \mathcal{S}$ for $k \in \mathbb{Z}_+$.
\item Agreement condition: 
There exists a finite time $k_a \geq 0$ such that 
$\text{Prob}\big\{x^N[k_a]\in \mathcal{C}_{n_N}\,|\; x[0]\bigr\}=1$, where
the consensus set $\mathcal{C}_{n_N}$ is defined as 
 \begin{equation*}\label{eqn: ConsensusSet}
 \mathcal{C}_{n_N}=\{ x \in \mathbb{Z}^ {n_N} |~x_1=\cdots=x_{n_N}\}.
 \end{equation*}
\end{enumerate}
\end{definition}

Next, we outline the algorithm employed for achieving consensus 
in the presence of misbehaving agents. The algorithm is the quantized 
version of the weighted mean subsequence reduced
(W-MSR) algorithm studied in \cite{DibIshiiIFAC,LeBlanc} and thus will 
be referred to as the QW-MSR algorithm. 
Using similar ideas, resilient consensus of second-order agent networks 
has been studied as well \cite{DibIshiiSCL,DibIsh:aut16}.

\begin{algorithm}[QW-MSR Algorithm]\label{alg:QW-MSR}
\begin{enumerate}
 \item At each time step $k$, if the normal agent $i$ makes an update
in its value, i.e., $i\in\mathcal{U}[k]$, then it 
receives the state values of its neighbors $j\in\mathcal{N}_i$ and 
sorts them in a descending order.
 \item If there are less than $f$ agents which have state values strictly larger than  $x_i [k]$, then the normal node $i$ ignores the incoming edges from those agents. Otherwise, it ignores the incoming edges from $f$ agents which have the largest state values. Similarly, if there are less than $f$ agents which have state values strictly smaller than $x_i [k]$, then node $i$ ignores all incoming edges from these nodes. Otherwise, it ignores the $f$ incoming 
edges from those which have the smallest values. 
	\item Apply the update rule \eqref{eqn: ProbQuanGenUpdatesNONDELAY} 
	by substituting $a_{ij}[k]=0$ for all edges $(j,i)$ which are neglected in step 2.
\end{enumerate}
\end{algorithm}

The main feature of this algorithm lies in its simplicity. 
Each normal node ignores the information received from its neighbors 
which may be misleading. 
\textcolor{black}{%
The normal agents do not make attempts
to identify the malicious agents in the network.
In particular, it always ignores
up to $f$ edges from neighbors whose values are large, and $f$ edges from neighbors whose values are small.} 
As we will see, in applying the algorithm, there is no need for information more than that of each agent's neighbors and the upper bound $f$ for the number of malicious agents.  
The underlying graph $\mathcal{G}[k]$ at time $k$ is determined by
the edges not ignored by the agents.
The adjacency matrix $A[k]$ and the Laplacian matrix $L[k]$ at time $k$
are determined accordingly. 

\textcolor{black}{%
The assumption on the number $f$ of malicious agents in the network
is standard in the problem setting of computer science (e.g., \cite{Lynch}). 
It will become clear that, for the MSR algorithm to function properly, 
the maximum $f$ depends on the size and the topology of the given network. 
There is a history of 
results showing the maximum number of malicious agents that can be 
tolerated in MSR-type algorithms when the network is 
a complete graph; see, for example, \cite{BenOr:83,Lynch,Rabin:83}.
Our main results are significant in that we obtain tight bounds on $f$ 
through the necessary and sufficient conditions
expressed in terms of robust graphs. }

The first problem studied in this paper is formulated as follows:
Under the $f$-total malicious model, find a condition on the network
topology such that the normal agents reach resilient quantized consensus 
almost surely using the QW-MSR algorithm outlined above.
We consider the case of synchronous updates and asynchronous updates
separately in Sections~\ref{Sect:synch} and \ref{Sect:Asynch}. 
Then, in Section~\ref{Sect: DELAY}, 
we study the more realistic situation when time delays are present in 
the communication among agents.


\section{Resilient Consensus for Synchronous Networks}\label{Sect:synch}

We provide the solution to the 
resilient consensus problem for the synchronous update case.
The result will be given in the form of a necessary and sufficient 
condition on the underlying network structure. 

\subsection{Characterization Based on Robust Graphs}

We are ready to present the main result of this section. The following theorem provides a necessary and sufficient condition 
for resilient quantized consensus for synchronous updating times. 
It shows that robustness in the network is a key property
to guarantee sufficient connectivity among the normal agents to avoid being 
misguided by the malicious agents. Let $\mathcal{S}$ be the interval given by 
\begin{equation}\label{eqn: SafetyNONDELAY}
  \mathcal{S}
    = \bigl[
        \min x^N[0],\max x^N[0]
      \bigr],
\end{equation}
where the minimum and maximum are taken over all entries of vectors. 
This set will be shown to be the safety interval. 

\begin{theorem}\label{Thm: NecSufSYNCHNONDELAY}\rm
Under the $f$-total malicious model, 
the network of quantized agents with the QW-MSR algorithm 
reaches resilient quantized consensus almost surely 
with respect to the randomized quantization
if and only if 
the underlying graph is $(f+1,f+1)$-robust. 
\end{theorem}

To establish quantized consensus in this probabilistic setting, 
we follow the arguments introduced in \cite{KashyapBasarSrikant}. 
In the following, we restate Theorem~2 from this reference 
with minor modifications to accommodate our problem setup. 

\begin{lemma}\label{lemma: Basisoftheproof}\rm
Consider the network of quantized agents interacting over 
the graph $\mathcal{G}$ through the QW-MSR algorithm. 
Suppose that the following three conditions are met for 
the normal agents:
\begin{enumerate}[(C1)]
\item There exists a set $\mathcal{S}$ such that for each normal agent $i$,
$x_i[k] \in \mathcal{S}$ for all $k\in\mathbb{Z}_+$ and all $x^N[0]$.
\item For each state $x[k]=x_0$ at time $k$, there exists a finite time $k_x$ such that $\text{Prob}\bigl\{x^N[k+k_x]\in\mathcal{C}_{n_N}\,|\,x[k]=x_0\bigr\} > 0$.
\item If $x[k]\in \mathcal{C}_{n_N}$, then $x[k'] \in \mathcal{C}_{n_N}$, $\forall k'>k$.
\end{enumerate}
Then, the network reaches quantized consensus almost surely. 
\end{lemma}

\begin{remark}\rm
The result in \cite{KashyapBasarSrikant} holds for a general class of 
algorithms, but there it is given for the case of quantized average consensus. 
In the lemma above, it is adapted to the regular quantized consensus. 
In the proof of Theorem~\ref{Thm: NecSufSYNCHNONDELAY}, we show that the QW-MSR algorithm satisfies the conditions (C1)--(C3) in the lemma when it is 
applied to robust graphs. Intuitively, if the algorithm satisfies these 
conditions for normal agents, then the scenarios for reaching 
consensus occur infinitely often with high probability. This is because the probability for such
an event to occur is positive based on the condition (C2). Then, as soon as normal agents reach consensus, they do not change their values by (C3).
\end{remark}

\textit{Proof of Theorem~\ref{Thm: NecSufSYNCHNONDELAY}}:
{(Necessity)} If the graph is not $(f+1,f+1)$-robust, the set of 
its nodes includes two disjoint and nonempty subsets $\mathcal{V}_1$ 
and $\mathcal{V}_2$ that do not meet any of the three conditions in 
Definition~\ref{Def: RobustGraph}. Thus, for $i=1,2,$ the total 
number of nodes in $\mathcal{V}_i$ that have at least $f+1$ 
incoming edges from $\mathcal{V}\setminus\mathcal{V}_i$ is 
less than $f+1$. 
Here, we take all malicious nodes to be in the sets $\mathcal{X}^{f+1}_{\mathcal{V}_1}$ 
and $\mathcal{X}^{f+1}_{\mathcal{V}_2}$.
It then follows that $\mathcal{V}_1 \setminus \mathcal{X}^{f+1}_{\mathcal{V}_1}$ 
and $\mathcal{V}_2 \setminus \mathcal{X}^{f+1}_{\mathcal{V}_2}$ 
are two disjoint and nonempty subsets of normal agents. 
Now, assign $a$, $b$ and $\lfloor {(a+b)/2 }\rfloor$ 
to the nodes in $\mathcal{V}_1$, $\mathcal{V}_2$, and the rest of the nodes, respectively, 
where $a,b\in\mathbb{Z}$ and
$a < b-1$. Suppose that the malicious agents do not change their state values. Then, the normal agents in $\mathcal{V}_1$ and $\mathcal{V}_2$ will ignore all of their neighbors that have different values from themselves and they will stay at their states. Thus, the normal agents 
contained in $\mathcal{V}_1 \setminus \mathcal{X}^{f+1}_{\mathcal{V}_1}$ and $\mathcal{V}_2 \setminus \mathcal{X}^{f+1}_{\mathcal{V}_2}$
remain at the values $a$ and $b$
at all times.
This implies that the agreement condition cannot be met.

{(Sufficiency)}
We must show that by applying QW-MSR to the network of $f$-total model, 
the conditions (C1)--(C3) in Lemma~\ref{lemma: Basisoftheproof} hold. 
First, we prove (C1), which is the safety condition.
Denote the minimum and maximum values of the normal agents at time $k$ by
\begin{equation}
\underline{x}[k] = \min x^N[k],~~
\overline{x}[k] = \max x^N[k].
\label{eqn:x_min_max}
\end{equation}
In the network, there are at most $f$ malicious agents, and, at each time step, 
each normal agent removes the values of at most $2f$ neighbors,
$f$ from above and $f$ from below. 
Hence, those faulty agents with values outside the
interval $\bigl[\underline{x}[k],\overline{x}[k]\bigr]$ are all ignored at each time step. 
In other words, each normal agent $i$ is affected by only the values within
$\bigl[\underline{x}[k],\overline{x}[k]\bigr]$. 
It thus follows that the value of the normal agent $i$ 
in the update rule \eqref{eqn: ProbQuanGenUpdatesNONDELAY} 
is upper bounded by
\begin{align*}
  x_i[k+1] 
    &
     \leq \Big\lceil 
            \sum_{j\in \mathcal{N}_i\cup\{i\}}  w_{ij}[k] x_j[k]
          \Big\rceil \\
    &\leq \Big\lceil 
           \sum_{j\in \mathcal{N}_i\cup\{i\}}  
               w_{ij}[k] \overline{x}[k]
          \Big\rceil
    = \big\lceil 
         \overline{x}[k]
      \big\rceil
    =\overline{x}[k].
\end{align*}
This implies that $\overline{x}[k+1] \leq \overline{x}[k]$, that is,
$\overline{x}[k]$ is a monotonically nonincreasing function of time. 
Likewise, agent $i$'s value can be bounded from below as
\begin{align*}
  x_i[k+1] 
   &
   \geq \Big\lfloor
          \sum_{j\in \mathcal{N}_i\cup\{i\}}
                 w_{ij}[k] x_j[k]
        \Big\rfloor \\
   &\geq \Big\lfloor 
           \sum_{j\in \mathcal{N}_i\cup\{i\}}
                 w_{ij}[k] \underline{x}[k]
         \Big\rfloor 
   = \big\lfloor 
        \underline{x}[k]
     \big\rfloor
   = \underline{x}[k].
\end{align*}
Thus, we have $\underline{x}[k+1] \geq \underline{x}[k]$, which shows that
$\underline{x}[k]$ is a monotonically nondecreasing function of time. 
Consequently, we conclude that for the normal agent $i$, its state satisfies
$x_i[k]\in\bigl[\underline{x}[k],\overline{x}[k]\bigr]\subset \mathcal{S}$ for all $k$ with the interval $\mathcal{S}$ given in \eqref{eqn: SafetyNONDELAY}. 
Thus, (C1) is established. 

Next, we prove (C2) in Lemma~\ref{lemma: Basisoftheproof}. 
Since $\underline{x}[k]$ and $\overline{x}[k]$ are contained in $\mathcal{S}$ and 
are monotone, there is a finite time $k_c$ such that they both reach their final values 
with probability 1. Denote the final values of $\underline{x}[k]$ and $\overline{x}[k]$ by $\underline{x}^*$ and $\overline{x}^*$, respectively. 
Now, to conduct a proof by contradiction, we assume $\underline{x}^* < \overline{x}^*$. 
Denote by $\mathcal{X}_1[k]$ the set of all agents including the malicious
ones at time $k \geq k_c$ with state values equal 
to $\overline{x}^*$ or larger. 
Likewise, denote by $\mathcal{X}_2[k]$ the set of agents 
whose states are equal to $\underline{x}^*$ or smaller. That is, 
\begin{equation}
\begin{split}
  \mathcal{X}_1[k] 
    &= \bigl\{i\in\mathcal{V}:~x_i[k]\geq \overline{x}^*\bigr\},\\
  \mathcal{X}_2[k] 
    &= \bigl\{i\in\mathcal{V}:~x_i[k]\leq \underline{x}^*\bigr\}.
\end{split}
\label{eqn:X}
\end{equation}
We show that with positive probability, the normal agents in $\mathcal{X}_1[k]$ 
decrease their values, and the normal agents in $\mathcal{X}_2[k]$ increase their values
at the next time step. 
These two sets $\mathcal{X}_1[k]$ and $\mathcal{X}_2[k]$ are nonempty and disjoint 
by assumption. Moreover, the underlying graph is $(f+1,f+1)$-robust.  
Thus, one of the conditions in Definition~\ref{Def: RobustGraph} must be fulfilled. 
In particular, there always exists a normal agent $i$ either in $\mathcal{X}_1[k]$ or 
$\mathcal{X}_2[k]$ with $(f+1)$ edges from $\mathcal{V}\setminus\mathcal{X}_1[k]$ or $\mathcal{V}\setminus\mathcal{X}_2[k]$, respectively. 
Without loss of generality, we suppose that the normal agent $i$ in $\mathcal{X}_1[k]$ has this property. Its value clearly is $x_i[k]=\overline{x}^*$.
In step~2 of QW-MSR, it neglects at most $f$ values from $\mathcal{V}\setminus\mathcal{X}_1[k]$. Hence, it makes an update using at least one agent from its neighbors whose value is smaller than $\overline{x}^*$. In step~2 of QW-MSR, it also neglects all values larger than $\overline{x}^*$ 
since there are at most $f$ such values coming from the malicious agents. 
Thus, by the update rule 
\eqref{eqn: ProbQuanGenUpdatesNONDELAY}, we write 
\begin{equation}\label{eqn: ShrinkingSets}
  x_i[k+1] 
   \leq Q\bigl( 
         (1-\alpha)\overline{x}^*+\alpha (\overline{x}^*-1) )
   = Q ( \overline{x}^*-\alpha ). 
\end{equation}
Here, by \eqref{eqn:Q},
the quantizer output takes the truncated value as 
$Q( \overline{x}^*-\alpha)= \overline{x}^*-1$ with probability $1-\alpha$. 
Thus, with positive probability, we have
\begin{equation*}
  x_i[k+1] \leq \overline{x}^* - 1.
\end{equation*}
This indicates that with positive probability, one of the normal agents taking 
the maximum value $\overline{x}^*$
decreases its value by at least one. 
Similarly, if the normal agent $i$ is in $\mathcal{X}_2[k]$, 
then with positive probability, it chooses the ceil quantization, in which case
its value will increase above $\underline{x}^*$.

We must next show that with positive probability, none of the normal 
agents in $\mathcal{V} \setminus \mathcal{X}_1[k]$ enters $\mathcal{X}_1[k+1]$
at the next time step.
If the normal agent $i$ at time $k$ is in $\mathcal{V} \setminus \mathcal{X}_1[k]$, it is upper bounded by $\overline{x}^*-1$. 
According to the update rule \eqref{eqn: ProbQuanGenUpdatesNONDELAY}, 
the inequality \eqref{eqn: ShrinkingSets} holds in this case as well. 
Thus, with probability $1-\alpha$, agent $i$ does not come into $\mathcal{X}_1[k+1]$. 
The same steps show that any of the normal agents in $\mathcal{V} \setminus \mathcal{X}_2[k]$ will not go into $\mathcal{X}_2[k+1]$ with positive probability. 

From the above, we conclude that for any $k \geq k_c +n_N$, 
the number of normal agents in one of the sets $\mathcal{X}_1[k]$ and $\mathcal{X}_2[k]$ 
is zero with positive probability because there are only $n_N$ such agents.
This is a contradiction and proves (C2).

In the last step, we have to show (C3) in Lemma~\ref{lemma: Basisoftheproof}, 
i.e., when all normal agents reach agreement
and go inside $\mathcal{C}_{n_N}$, they stay there from that time on. 
Assume that the normal agents have reached the common value $x^*$. 
Since the maximum number of malicious agents is $f$, all such agents $j$ with 
values $x_j[k] \neq x^*$ are ignored by the normal agents. Thus, in the third step of QW-MSR, when they apply the update rule \eqref{eqn: ProbQuanGenUpdatesNONDELAY}, $x_i[k+1]=x^*, \forall i \in \mathcal{V} \setminus \mathcal{M}$. We have thus shown (C3), and this concludes the proof.
\hfill \mbox{$\blacksquare$}

The $(f+1,f+1)$-robustness as a necessary and sufficient condition is consistent with 
the resilient consensus problems in \cite{LeBlanc} and 
\cite{DibIsh:aut16} for 
the real-valued agent cases with first-order and second-order dynamics, respectively. 
However, these works consider consensus in real-valued agent networks and 
establish convergence in an asymptotic fashion; moreover, the updating rules there
are without any randomization. 
This paper studies agents taking 
quantized values and the convergence is in finite time in a probabilistic sense. 

It is noted that our approach can be applied to the binary-valued consensus
case \cite{BenOr:83,FisLynPat:85,Lynch,Rabin:83}.
As long as the initial states of all agents are restricted to 
0 and 1, the safety interval in \eqref{eqn: SafetyNONDELAY} indicates
that the normal agents' values will remain binary and come to agreement
eventually. This fact remains true for all results presented in this paper. 

\begin{remark}\label{remark:conv_time}\rm
Analysis on the convergence rate for quantized consensus problems 
has recently gained attention, where upper bounds on the worst-case
convergence times are derived with respect to the number $n$ of
agents. In \cite{KashyapBasarSrikant}, some common classes of agent 
networks are considered while 
further studies on static and time-varying networks can be found in 
\cite{EstBas:16,ZhuMar:11}. 
It is however noted that these works deal with the problem of
average consensus where the underlying graph is undirected.
The paper \cite{CaiIsh:12} carries out an analysis for
average quantized consensus over directed graphs based on a
specific algorithm employing additional dynamics in each agent. 
In contrast to these works, in our resilient consensus problem,
the algorithm is synchronous, but the network is directed
and moreover switches according to the MSR mechanism. 
Hence, for the convergence time analysis of the algorithm, 
it seems difficult to apply existing results. This problem 
is left for future research.  
\end{remark}

We now consider the special case when no malicious agent is present in
the network (i.e., with $f=0$). Then, the QW-MSR algorithm reduces to the update rule 
in \eqref{eqn: ProbQuanGenUpdatesNONDELAY} with a static matrix $W$.
The following corollary of Theorem~\ref{Thm: NecSufSYNCHNONDELAY}
demonstrates that in this case, to achieve quantized consensus, 
it is necessary and sufficient that a spanning tree exists in the network, 
which is the well-known topological condition 
for multi-agent consensus when the states take real values \cite{mesbahi}. 

\begin{corollary}\rm\label{Corollory: ShorterProofForNormalNetworkNONDELAY}
When no malicious agent is present, 
the network of quantized agents based on the update rule \eqref{eqn: ProbQuanGenUpdatesNONDELAY} 
reaches quantized consensus almost surely 
with respect to the randomized quantization
if and only if the underlying graph has a directed spanning tree.
\end{corollary}

\textit{Proof}:\ With $f=0$, by Theorem~\ref{Thm: NecSufSYNCHNONDELAY}, 
the necessary and sufficient condition for reaching consensus in normal networks is $(1,1)$-robustness. Then, by Lemma~\ref{lemma: robustgraphs}~(iii), a graph is $(1,1)$-robust, or 1-robust, if and only if it has a  directed spanning tree.
\hfill \mbox{$\blacksquare$}

In the literature of quantized consensus, 
to the best of our knowledge, 
the above necessary and sufficient condition has not
been reported elsewhere. 
Related works include \cite{NedOlshOzdTsitsilis} which
established strong connectivity as a sufficient condition,
and \cite{LavaeiMurray,KashyapBasarSrikant} which deal with
undirected communications. In contrast, we have studied
the more general directed graphs. 
The same condition appears in \cite{CaiIsh:12}, but 
is for a specific class of gossip-based time-varying networks. 
In Section~\ref{Sect: DELAY},
we will present further generalization for the case
with delays in communication. 


\subsection{Role of Probabilistic Quantization}

Quantization is necessary in the update scheme \eqref{eqn: ProbQuanGenUpdatesNONDELAY} 
of the agents for keeping their states to take integer values from
the weighted average of the neighbors' states.
In this subsection, we show that randomization in the quantizers 
plays an important role in the consensus problem. 
Similar discussions can be found in, e.g., \cite{Aysal,CarliFagnaniFrascaZampieri,ChamieLiuBasar,KarMoura,KashyapBasarSrikant}, 
but are focused on average consensus over undirected graphs without any malicious agents.

\begin{figure}[t]
 	\centering
 	\vspace*{3mm}
 	\includegraphics[width=34mm]{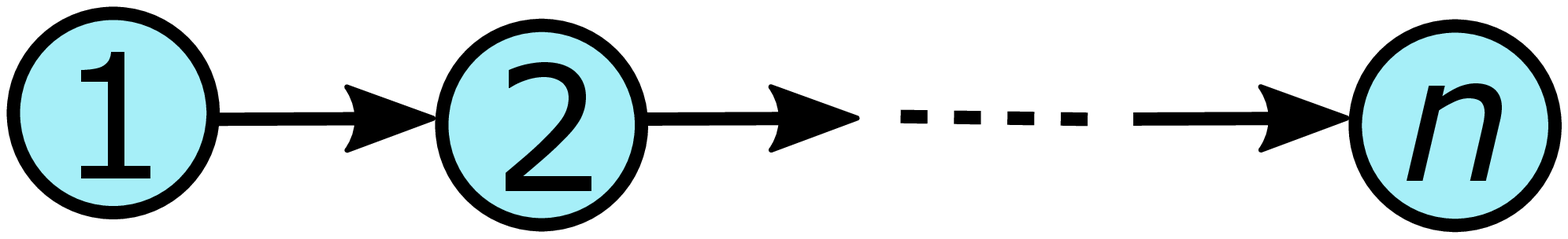}
 	\vspace*{1mm}
 	\caption{A line graph with no malicious agent fails to reach consensus 
with ceil quantizers when the initial value for node $i$ is taken to be
its index $i$.}
 	\label{fig: linegraph}
	\vspace*{8mm}
	\includegraphics[width=28mm]{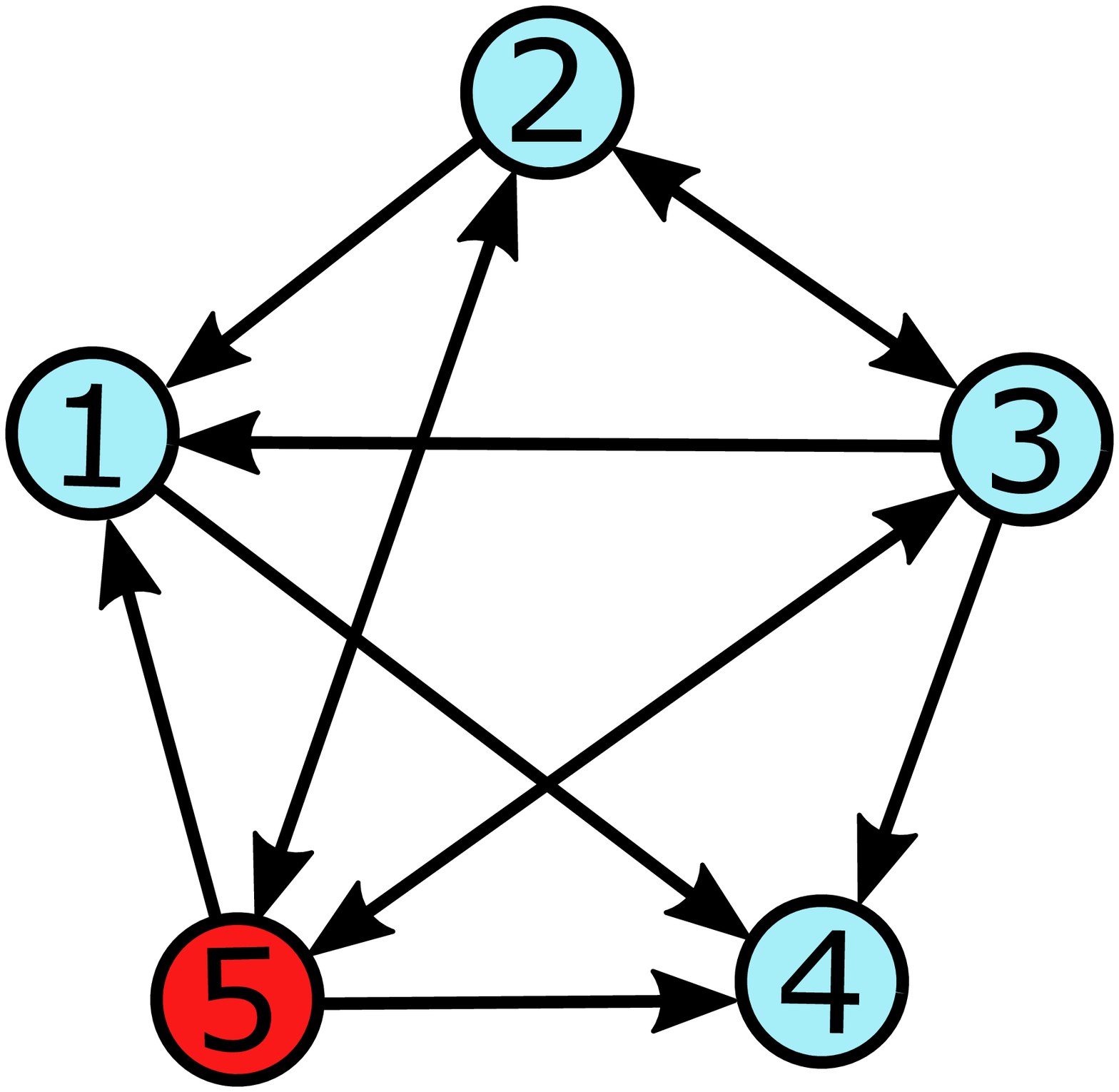}
	\vspace*{1mm}
	\caption{A $(2,2)$-robust graph which fails to reach consensus by using 
         ceiling quantization or floor quantization.}
	\label{fig: 5node22robust}
\end{figure}

First, to show the limitation of deterministic quantization even 
when no malicious agent is present in the network (i.e., $f=0$), 
we use the line graph example in Fig.~\ref{fig: linegraph}.
Suppose that in the update rules, the ceil quantizers $Q(y)=\ceil y$ are used
instead of the probabilistic ones. 
Let agent~$i$ take an initial value of $x_i[0]=i$ for all $i$.
Then, we easily conclude that 
the agents stay at their initial states for all times
and thus consensus is impossible. A similar argument holds for the
floor quantizer  $Q(y)=\floor y$ by changing the initial values to $x_i[0]=n-i$.
Note however that line graphs contain spanning trees.
It is clear that neither ceil nor floor is sufficient
in such examples and we need a combination of both with a suitable switching mechanism. 
This can be achieved by means of the probabilistic quantizers in \eqref{eqn:Q}.

Next, we provide an example with a malicious agent. 
The network in Fig.~\ref{fig: 5node22robust} with five nodes is $(2,2)$-robust.
We set $f=1$ and take agent~$5$ to be malicious. This agent keeps its value 
unchanged from the initial time.
The values of the agents are initialized as $x[0]=[2~2~2~3~5]^T$.
Using ceil quantizers in QW-MSR forces each normal agent to 
stay at its initial value, and hence no consensus is reached. 
In this case, the malicious agent~5 is ignored
at all times since it is set to be the largest value. 
It is interesting to notice that removal of node~5 from the graph in 
Fig.~\ref{fig: 5node22robust} results in a graph with a spanning tree.
We can reach similar conclusions with floor quantization
by initializing the values as $x[0]=[4~4~4~3~1]^T$.

At a more general level, we can explain the importance of the probabilistic
quantization in the proof of Theorem~\ref{Thm: NecSufSYNCHNONDELAY} as follows.
There, two disjoint and nonempty subsets $\mathcal{X}_1[k]$ and $\mathcal{X}_2[k]$ 
have been introduced. It is shown that one of them loses all normal agents in 
finite time steps almost surely. 
In one of these sets, there is a normal agent $i$ 
that has at least $f+1$ edges from nodes outside the set it belongs to. 
Now, if agent $i$ is in the set $\mathcal{X}_1[k]$ and if the quantization is 
always based on ceiling, there is no guarantee that it goes out of $\mathcal{X}_1[k+1]$. 
Likewise, if agent $i$ is contained in $\mathcal{X}_2[k]$ and if floor
quantization is employed at all times, then 
this agent will remain in $\mathcal{X}_2[k+1]$.


\section{Asynchronous Networks: Enhancing Resilience
via Probabilistic Updates}
\label{Sect:Asynch}

In this section, we consider asynchronous update schemes. 
We will highlight that asynchrony in the updates can create 
weaknesses in the algorithm that can be exploited by the malicious
agents. However, we show that this problem can be overcome by employing
a probabilistic updating scheme.


\subsection{Deterministic Update Scheme}

Recall that the QW-MSR algorithm in Algorithm~\ref{alg:QW-MSR}
is applicable to the case of asynchronous update rules. 
The set of normal agents updating at time $k$ is represented
by $\mathcal{U}[k]$. When no update is made, 
the control input simply becomes $u_i[k]=0$ and thus the state remains
unchanged as $x_i[k+1]=x_i[k]$. 

In the deterministic setting, we assume that
each normal agent $i$ makes an update at least once 
in $\bar{k}$ time steps, that is,
\begin{equation}
  \bigcup\limits_{\ell=k}^{k+\overline{k}-1}  
      \mathcal{U}[\ell]=\mathcal{V}\setminus\mathcal{M}~~
  \text{for $k\in\mathbb{Z}_+$}.
\label{eqn:det_update}
\end{equation}
On the other hand, the malicious agents need not 
follow these schemes and may even be aware of the updates of the whole 
network.

Now, we state a sufficient condition for deterministic asynchronous 
networks in terms of graph robustness, but with a more restrictive 
requirement than that in Theorem~\ref{Thm: NecSufSYNCHNONDELAY}.

\begin{theorem}\rm\label{Thm: SufCondDeterminiticAsynchNONDELAY}
Under the $f$-total malicious model, 
the network of asynchronous quantized agents 
with the QW-MSR algorithm satisfying \eqref{eqn:det_update} 
reaches resilient quantized consensus almost surely 
with respect to the randomized quantization
if the underlying graph is $(2f+1)$-robust.
\end{theorem}

The proof of Theorem~\ref{Thm: SufCondDeterminiticAsynchNONDELAY} 
is skipped for now. In the next section, we present 
a more general result for deterministic asynchronous networks,
from which the theorem above 
follows (see Theorem~\ref{Thm: SufCondDeterDELAY} and Remark~\ref{remark:delay}).

It is noted that a $(2f+1)$-robust graph is also $(f+1,f+1)$-robust 
by Lemma~\ref{lemma: robustgraphs}~(ii).  This indicates that there is
a gap between Theorem~\ref{Thm: NecSufSYNCHNONDELAY} and 
Theorem~\ref{Thm: SufCondDeterminiticAsynchNONDELAY} for 
the synchronous scheme and the asynchronous scheme, respectively. 
This gap originates from the fact that asynchrony in the updating times 
creates more ways for malicious agents to deceive the normal agents. 
This point is demonstrated through an example in the next subsection.

\subsection{Discussion on Synchronous versus Asynchronous}
\label{sec:discussion1}

\begin{figure}[t]
	\centering
	\vspace*{4mm}
	\includegraphics[width=28mm]{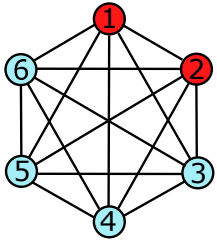}
	\vspace*{1mm}
	\caption{A complete graph which is $(3,3)$-robust, but fails to reach consensus under asynchronous updates when agents 1 and 2 are malicious.}
	\label{fig: 6nodeasynch}
\end{figure}

Now, we discuss why the sufficient condition for the case
of asynchronous normal agents is more restrictive than the
case of synchronous agents in the previous section.

Consider the 6-node complete graph depicted in 
Fig.~\ref{fig: 6nodeasynch}. This graph is $(3,3)$-robust 
due to Lemma \ref{lemma: robustgraphs} (iv). 
In this graph, agents~1 and 2 are taken to be malicious. 
They switch their values according to  $x_i[2m]=i$ and 
$x_i[2m+1]=i+6$ for $m \in \mathbb{Z}_+$. 
For each agent $i$, let's assign $x_i[0]=i$. 
Also, assume that agents~3 and 4 make updates only at even time 
steps ($k=2m$) and agents~5 and 6 make updates only at odd time 
steps ($k=2m+1$). By applying the QW-MSR algorithm, 
we observe that with probability 1, agents 3 and 4 reach 
agreement among themselves (at either 3 or 4), and agents~5 and 6
agree upon a different value (at either 5 or 6). 
Here, although the underlying graph is complete, 
the agents cannot reach consensus because the graph is 
not sufficiently robust to have two malicious agents.

Using Lemma~\ref{lemma: robustgraphs} (iv) and the above results, 
we note that, with less than $4f+1$ nodes in the network, it is impossible to reach 
consensus in $f$-total malicious models under asynchronous updates 
even when the graph is complete;
there will be some scenarios of updates and adversaries' behaviors that prevent it. 
However, when the normal agents make updates synchronously, 
having a complete graph with $2f+1$ nodes is sufficient to reach consensus.

We would like to tighten the topological condition 
for asynchronous updates.
As seen in the above example, since the normal agents~3 
and 4 make updates at time steps 
when the normal agents 5 and 6 are in idle mode, the malicious agents take advantage of 
this asynchrony and appear at different states at the times of their updates. 
For instance, at the time of updates, agents 3 and 4 receive values $x_1[k]=1$ and 
$x_2[k]=2$ from the malicious agents 1 and 2. Then, the values from agents 1, 2, 5, and 6 
are ignored.

Such an undesirable situation can be avoided, for example, 
if there are chances that all normal agents simultaneously make updates,
even if they occur very seldom. 
This type of feature in fact will allow us to obtain 
$(f+1,f+1)$-robustness as a necessary and sufficient 
condition for resilient consensus. 
To this aim, we make use of randomized update times.

\subsection{Probabilistic Update Scheme}\label{subsec:prob}

In the probabilistic setting, we assume that each normal 
agent~$i$ makes an update 
at time $k \geq 0$ with probability $p_i \in(0,1]$ 
in an i.i.d.\ fashion. 
That is, for agent~$i$, at each time $k$,
\begin{equation}
\begin{split}
&\text{Prob}\bigl\{i\in\mathcal{U}[k]\bigr\} = p_i,~~
\text{Prob}\bigl\{i\notin\mathcal{U}[k]\bigr\} = 1-p_i.
\label{eqn:prob_edge}
\end{split}
\end{equation}
Note that with such updates, the algorithm remains fully 
distributed. Even the probabilities $p_i$ need not be identical. 


An advantage of introducing randomization for the normal agents 
is that the malicious agents cannot predict the update times 
in advance.
In this respect, randomization in update times is utilized 
as a defensive means against potential conspiracy of the 
malicious agents. 
Moreover, there is always nonzero probability that any normal 
agents in the system update their states. This feature enables 
us to establish a stronger result than that for the deterministic
case. 

The following theorem is the main result of this section. 
It shows that for the probabilistic updating scheme, the requirement
on graph robustness is the same as that 
for the synchronous case in Theorem~\ref{Thm: NecSufSYNCHNONDELAY}.

\begin{theorem}\rm\label{Thm: NecSufCondRandomAsynchNONDELAY}
Under the $f$-total malicious model, 
the network of randomly asynchronous quantized agents 
with the QW-MSR algorithm satisfying \eqref{eqn:prob_edge} 
reaches resilient quantized consensus almost surely 
with respect to the randomized quantization and updates
if and only if the underlying graph is $(f+1,f+1)$-robust. 
\end{theorem}

\textit{Proof}:
	{(Necessity)}
	This part follows from the necessity part of 
Theorem~\ref{Thm: NecSufSYNCHNONDELAY}
	since the synchronous updating scheme is a special case of 
	asynchronous updating schemes. 
	
	{(Sufficiency)}
	We must show that the three conditions (C1)--(C3) in 
Lemma~\ref{lemma: Basisoftheproof} hold. 
This can be done as in the proof of 
Theorem~\ref{Thm: NecSufSYNCHNONDELAY}, and thus we provide only an outline for the probabilistic 
	asynchronous case considered here. 
	
	For (C1), it is enough to notice that
	the values $\underline{x}[k]$ and $\overline{x}[k]$ as defined 
	in \eqref{eqn:x_min_max} are bounded in the interval $\mathcal{S}$
	in \eqref{eqn: SafetyNONDELAY} and are monotone. 
	Furthermore, these facts imply that 
	$\underline{x}[k]$ and $\overline{x}[k]$ arrive at their
	final values $\underline{x}^*$ and $\overline{x}^*$ in some
	finite time $k_c$ almost surely. 
	
	We show (C2) by contradiction and assume $\underline{x}^*<\overline{x}^*$. 
	Take the sets $\mathcal{X}_1[k]$ and $\mathcal{X}_2[k]$ as in \eqref{eqn:X}.
	By assumption, these sets are disjoint and nonempty.
	We have two claims at this point.
	The first is that, at each $k \geq k_c$, a normal agent in 
	at least one of these sets goes out from the corresponding set 
	at the next time step with positive probability. 
	This can be shown because 
	by $(f+1,f+1)$-robustness, there is a normal agent $i$ 
	in either $\mathcal{X}_1[k]$ or $\mathcal{X}_2[k]$ which has
	at least $f+1$ incoming links from outside the set
	$\mathcal{X}_1[k]$ or $\mathcal{X}_2[k]$, respectively.
	In the probabilistic
	asynchronous updating case, for this normal agent $i$,
	the probability to update at this time $k$ is positive. 
	The second claim is that with positive probability, none of the normal agents outside  $\mathcal{X}_j[k]$ enters $\mathcal{X}_j[k+1]$ for $j=1,2$. This can be proved similarly
	as in the first one. 
	By these two claims, we conclude that
	for any $k\geq k_c + n_N$, one of the two sets,
	$\mathcal{X}_1[k]$ or $\mathcal{X}_2[k]$, contains no normal agent with positive probability, which is clearly a contradiction. 
	
	The condition (C3) holds as well since once all of the normal agents arrive 
	at consensus, whether a normal agent makes an update or not, it will 
	keep its state unchanged at the consensus value. 
\hfill \mbox{$\blacksquare$}


\section{Resilience under Delayed Communications}\label{Sect: DELAY}

Thus far, we assumed that the interactions among the agents
do not experience any time delay and the neighbors' information
can be transferred immediately. 
In this section, we consider the more realistic situation where
agents have access to only delayed information of the 
neighbors' states. We demonstrate that the malicious agents 
can exploit the delays to prevent consensus among the normal
agents when the graph is not sufficiently robust. 

\subsection{Problem Formulation and Algorithm}

In the case with delayed information, the control input
of the normal agent $i$ in \eqref{eqn:u_i} is given by
 \begin{equation}
 u_i[k] =
 Q\biggl(\sum_{j\in \mathcal{N}_i} 
       a_{ij}[k] (x_j[k-\tau_{ij}[k]]-x_i[k])
   \biggr),
 \label{eqn: ProbQuanGenControlDELAY}
\end{equation}
where the delays $\tau_{ij}[k]$ are present in each 
communication channel. They are assumed to be time varying, 
non-uniform, and bounded, i.e., $0\leq \tau_{ij}[k] \leq \overline{\tau}$ 
for some nonnegative constant $\overline{\tau}$. Note that this bound
$\overline{\tau}$ need not be known to any of the normal agents. 
As before, non-updating agents apply no input, i.e., $u_i[k]=0$ with
$a_{ij}[k]=0$ for $j\in\mathcal{N}_i$ if $i\notin\mathcal{U}[k]$.

For the malicious agent, in contrast, we assume that the
control input $u_i[k]$ can be arbitrary and may be
based on the current (non-delayed) state information
of any agent in the network. 

To ease the problem formulation, we introduce the following 
notations.
Let $D[k]$ be a diagonal matrix whose $i$th entry is given by
\begin{equation} \label{eqn: Dmatrix}
d_{i}[k]=\sum_{j=1}^{n} a_{ij}[k].
\end{equation} 
Then, let the matrices $A_{\ell}[k]\in\mathbb{R}^{n\times n}$ 
for $ 0 \leq \ell \leq \overline{\tau}$, 
and $L_{\overline{\tau}}[k]\in\mathbb{R}^{n\times(\overline{\tau}+1) n}$, 
respectively, be given by 
\begin{align}\label{eqn: A_lMatrices}
A_{\ell}[k]
 = \begin{cases} 
     a_{ij}[k] & \text{if $i\neq j$ and $\tau_{ij}[k]=\ell$},\\
     0 & \text{otherwise},
   \end{cases}
\end{align}
and 
\begin{equation*}
 L_{\overline{\tau}}[k]
    = \begin{bmatrix}
        D[k]-A_0[k] & -A_1[k] & \cdots & -A_{\overline{\tau}}[k]
      \end{bmatrix}.
\end{equation*}
Furthermore, let
\begin{equation*}
 W_{\overline{\tau}}[k]
  = \begin{bmatrix}
       I_{n}  & 0 & \cdots & 0
    \end{bmatrix}
     - L_{\overline{\tau}}[k].
\end{equation*}
Similarly to the usual Laplacian matrices, $L_{\overline{\tau}}[k]$ 
has the property that the sum of elements of each row is zero.  
Thus, each row sum of $W_{\overline{\tau}}[k]$ is one, and 
each non-zero entry of $W_{\overline{\tau}}[k]$ is lower bounded by some 
$\beta\in(0,1)$. 

Finally, denote by $z[k]\in\mathbb{Z}^{(\overline{\tau}+1)n}$ 
the extended state vector containing the past states given by
\begin{equation*}
 z[k]
  = \begin{bmatrix} 
      x^T[k] & x^T[k-1] & \cdots & x^T[k-\overline{\tau}]
    \end{bmatrix}^T.
\end{equation*}

\textcolor{black}{%
The update rule for each agent $i$ is based on 
the control \eqref{eqn: ProbQuanGenControlDELAY} for a normal
agent. 
Meanwhile, the malicious agents can choose their control arbitrarily 
and thus for them, we leave $u_i[k]$ unspecified at the current stage. 
Consequently, the update schemes for the agents can be written as }
\begin{equation}\label{eqn: ProbQuanGenUpdateDELAY}
  x_i[k+1]
   = \begin{cases}
       Q(e_i^T W_{\overline{\tau}}[k] z[k]) 
          & \text{if $i\in\mathcal{V}\setminus\mathcal{M}$},\\
       x_i[k] + u_i[k] & \text{if $i\in\mathcal{M}$},
     \end{cases}
\end{equation}
where $e_i\in\mathbb{R}^{(\overline{\tau}+1)n}$ is the unit vector
whose $i$th entry is 1 and the rest are zero.
Thus, each agent makes an update by a quantized convex combination of 
its neighbors' state values which may be delayed. 

The objective of the agent network is to reach resilient consensus among 
all agents by applying the local control rule \eqref{eqn: ProbQuanGenControlDELAY} 
with the probabilistic quantizers under the deterministic update scheme 
\eqref{eqn:det_update}. To this end, we again employ the QW-MSR algorithm
in Algorithm~\ref{alg:QW-MSR},
but this time, we interpret the neighbors' 
states to be the delayed information available at each update time $k$.

From the viewpoint of multi-agent systems in control,
the delay model in the input \eqref{eqn: ProbQuanGenControlDELAY} is 
reasonable and commonly adopted \cite{RenCao:11}. It uses the
most recent information of the neighbors available at the time
of updates. In contrast, the works \cite{AzaKie:02,Vaidya} from the computer science 
literature employ a delay model based on the so-called \textit{rounds}. 
An agent transmits its value along with its current round, i.e., 
the number of transmissions that it made so far. Updates are made
at an agent only after it receives the neighbors' data with the same round. 
This indicates that if there is one agent whose transmission takes
time, all of its neighbors would have to wait before making the next update,
which clearly slows down the convergence of the overall system.

\subsection{Robust Graph Conditions: Deterministic Updates}

We are now in the position to state the main result of this section. 
The theorem below establishes a sufficient condition 
for resilient quantized consensus in the asynchronous 
setting with delays and deterministic updating times. 
The randomized counterpart will be given later.

Let $\mathcal{S}_{\overline{\tau}}$ be the interval given by 
\begin{equation}
  \mathcal{S}_{\overline{\tau}}
    = \bigl[
        \min z^N[0],\max z^N[0]
       \bigr],
\label{eqn: SafetyDELAY}
\end{equation}
where the minimum and maximum are taken over all entries of the vector
$z^N[0]$ containing the initial values of normal agents. 
This set will be shown to be the safety interval. 

\begin{theorem}\label{Thm: SufCondDeterDELAY}\rm
Under the $f$-total malicious model, 
the network of quantized agents with delayed information and 
deterministic asynchronous update times 
in the asynchronous QW-MSR algorithm 
reaches resilient 
quantized consensus almost surely 
with respect to the randomized quantization
if the underlying graph is $(2f+1)$-robust. Moreover, the safety interval is determined by $\mathcal{S}_{\overline{\tau}}$ in \eqref{eqn: SafetyDELAY}.
\end{theorem}

\textcolor{black}{%
The proof of this theorem is presented in the Appendix. }

\begin{remark}\label{remark:delay}\rm
Theorem~\ref{Thm: SufCondDeterDELAY} is a generalized version of
Theorem~\ref{Thm: SufCondDeterminiticAsynchNONDELAY} 
where no delay is assumed with $\overline{\tau}=0$. 
The proof of Theorem~\ref{Thm: SufCondDeterDELAY} 
may appear similar to that of Theorem~\ref{Thm: NecSufSYNCHNONDELAY} 
for the synchronous problem. However, there are considerable 
differences between them. 
One technical difference 
lies between the definitions of the two sets $\mathcal{X}_{1\overline{\tau}}[k]$ and 
$\mathcal{X}_{2\overline{\tau}}[k]$ 
and those of $\mathcal{X}_1[k]$ 
and $\mathcal{X}_2[k]$ in Section~\ref{Sect:synch}. 
As a consequence, further discussions are required for showing that 
$\mathcal{X}_{1\overline{\tau}}[k]$ and $\mathcal{X}_{2\overline{\tau}}[k]$ 
are nonempty for $k \geq k_c$, which is not the case 
in the proof of Theorem~\ref{Thm: NecSufSYNCHNONDELAY}.
Notice that the sets $\mathcal{X}_{1\overline{\tau}}[k]$ and 
$\mathcal{X}_{2\overline{\tau}}[k]$ do not include the malicious agents, 
while the sets $\mathcal{X}_1[k]$ and $\mathcal{X}_2[k]$ there 
involve both normal and malicious agents. 
This difference originates from the definitions of 
$(2f +1)$- and $(f+1,f+1)$-robust graphs. In fact, 
in the $f$-total model, we see that the second term 
$f+1$ in $(f+1,f+1)$ guarantees that at least one of 
the agents in $\mathcal{X}_1[k]$ or $\mathcal{X}_2[k]$ 
is normal and has a sufficient number of incoming links 
for convergence. However, $(2f+1)$-robustness is a more 
local notion. Since the worst-case behavior of the 
malicious agents happens in the neighborhood of 
each normal agent, the sets $\mathcal{X}_{1\overline{\tau}}[k]$
and $\mathcal{X}_{2\overline{\tau}}[k]$ are defined in this way. 
\end{remark}

Note that the networks with asynchrony and delays are 
the generalized case of the synchronous networks without time delays. 
Thus, for the necessary condition on the network structure, 
the result of Theorem~\ref{Thm: NecSufSYNCHNONDELAY} is 
valid here. This fact is stated as a proposition in the following.

\begin{proposition}\label{Thm: NecCond}\rm
Under the $f$-total malicious models, 
if the network of quantized agents with delayed information 
in the QW-MSR algorithm
reaches resilient quantized consensus almost surely 
with respect to the randomized quantization,
then the underlying graph is $(f+1,f+1)$-robust. 
\end{proposition}

\subsection{Effect of Non-uniform Delays and Probabilistic Updates}

Similarly to the discussion in Section~\ref{sec:discussion1}, 
for networks with delays, there is also a gap between 
the sufficient condition in Theorem~\ref{Thm: SufCondDeterDELAY}
and the necessary condition in Proposition~\ref{Thm: NecCond}.
%
We have seen in Section~\ref{subsec:prob} that 
using randomization can be effective to obtain tighter results.
In particular, randomization enables the normal agents to 
have a chance to make updates at the same time. 
This has facilitated us to obtain $(f+1,f+1)$-robustness 
as a necessary and sufficient condition for resilient consensus 
of asynchronous networks without delay. 

However, when there are delays in the network, it seems 
difficult to gain similar benefits through randomization.
By exploiting non-uniform delays, 
a malicious agent can appear as having different states
at the same time by different normal agents. 
This is possible if the agent changes its value, 
but sends its state with different delays.
This means that randomization in update times may not be enough 
in such cases to relax the topological condition. 
We will examine this type of malicious behavior in the 
numerical examples in the next section. 

The following proposition gives a sufficient condition
with the same condition in terms of graph robustness
as in the deterministic case in Theorem~\ref{Thm: SufCondDeterDELAY}.

\begin{proposition}\rm\label{Prop: NecSufCondRandomAsynchDELAY}
Under the $f$-total malicious model, 
the network of randomly asynchronous quantized agents 
with time-varying delayed information satisfying \eqref{eqn:prob_edge} 
in the QW-MSR algorithm 
reaches resilient quantized consensus almost surely 
with respect to the randomized quantization and updates
if the underlying graph 
is $(2f+1)$-robust. Moreover, the safety interval is 
determined by $\mathcal{S}_{\overline{\tau}}$ in \eqref{eqn: SafetyDELAY}.
\end{proposition}

\textit{Proof}:
We must show that the three conditions (C1)--(C3) in 
Lemma~\ref{lemma: Basisoftheproof} hold. This can be done 
as in the proof of Theorem~\ref{Thm: SufCondDeterDELAY}. 
For the randomized 
updates considered here, the only difference is in showing (C2). 
In the deterministic scheme, each normal agent makes 
updates at least once in each $\bar{k}$ whereas in the 
probabilistic scheme the agents make updates at each 
time step with probability $p_i>0$. This fact reflects 
on the minimum $\bar{k}\cdot n_N$ and $n_N$ steps 
for $\mathcal{X}_{1\overline{\tau}}[k_c]$ and $\mathcal{X}_{2\overline{\tau}}[k_c]$ 
to become empty with positive probability in the 
deterministic and probabilistic schemes, respectively.  
\hfill \mbox{$\blacksquare$}

The $(2f+1)$-robustness as a sufficient condition is consistent with the resilient consensus problems in \cite{DibIshiiIFAC} and \cite{DibIsh:aut16} for the real-valued agent cases with first-order and second-order dynamics, respectively. 
However, these works consider 
consensus 
without any randomization in their updates. 
This paper studies agents taking only quantized values and the convergence 
is in finite time in a probabilistic sense.

\subsection{Further Results}

In what follows, we discuss a few extensions of our results
concerning agent systems with communication delays. 

\subsubsection{Normal Networks}
We consider the special case when no malicious agent is 
present in the network with $f=0$. Then, the QW-MSR algorithm 
reduces to the update rule in \eqref{eqn: ProbQuanGenUpdateDELAY} 
with a static matrix $W_{\overline{\tau}}$.
The following corollary of Theorem~\ref{Thm: SufCondDeterDELAY}
demonstrates that to achieve quantized consensus in this 
setting with delays and asynchrony, it is necessary and sufficient 
that a spanning tree exists in the network. This is the well-known 
topological condition for multi-agent consensus when the states 
take real values \cite{mesbahi}. 

\begin{corollary}\rm\label{Corollory: ShorterProofForNormalNetworkDELAY}
When no malicious agent is present, the network of quantized agents 
based on the update rule \eqref{eqn: ProbQuanGenUpdateDELAY} with 
delayed information reaches quantized consensus almost surely 
with respect to the randomized quantization
if and only if the underlying graph has a directed spanning tree.
\end{corollary}

\textit{Proof:}
With $f=0$, the conditions in Theorem~\ref{Thm: SufCondDeterDELAY} and 
Proposition~\ref{Thm: NecCond} coincide. Consequently, we obtain a necessary 
and sufficient condition for reaching quantized consensus in normal
networks to be $(1,1)$-robust. By Lemma~\ref{lemma: robustgraphs} (iii), 
this condition is equivalent to having 
a directed spanning tree.\hfill \mbox{$\blacksquare$}


The above corollary is an extension of 
Corollary~\ref{Corollory: ShorterProofForNormalNetworkNONDELAY} 
where there is no information delay in the network. 
The system model in this section with delays 
also generalizes the models studied for quantized 
consensus in \cite{Aysal,CaiIshii,CarliFagnaniFrascaZampieri,
ChamieLiuBasar,FrascaZampieri,GravelleMartinez,Li}. 


\subsubsection{$f$-Local Malicious Models}
The following corollary states that the same condition fulfills the sufficiency of $f$-local malicious models instead of the $f$-total models studied so far. This is because in the proofs of Theorem~\ref{Thm: SufCondDeterDELAY} and 
Proposition~\ref{Prop: NecSufCondRandomAsynchDELAY}, the presence of at most $f$ malicious agents in the neighborhood of each normal agent is the only assumption. In other words, the total number of malicious agents is not used in the proof. Note that the necessary condition stated in Proposition~\ref{Thm: NecCond} is valid for $f$-local case as well.

\begin{corollary}\rm\label{Corollory: f-localDeterministicSufCond}
Under the $f$-local malicious models, 
the network of deterministic/randomized asynchronous quantized agents 
with/without information delays in the asynchronous QW-MSR algorithm 
reaches resilient quantized consensus almost surely with respect to
the randomized quantization and updates
if the underlying graph is $(2f+1)$-robust
and only if the graph is $(f+1,f+1)$-robust.
\end{corollary}

Finally, we mention that in the results of Theorem~\ref{Thm: SufCondDeterDELAY}, 
Proposition~\ref{Prop: NecSufCondRandomAsynchDELAY}, and Corollary~\ref{Corollory: f-localDeterministicSufCond}, the minimum number of nodes in the underlying graph 
to reach resilient quantized consensus with asynchronous QW-MSR algorithms over 
$f$-total/$f$-local malicious models is $4f+1$ (with complete graphs). 
This minimum originates from Lemma~\ref{lemma: robustgraphs} (iv).
It is noteworthy that for real-valued first-order asynchronous settings, 
the sufficient condition $(3f+1)$-robustness proposed in \cite{LeBlancPhD} 
has been improved in this paper and in \cite{DibIshiiIFAC}. 
For a thorough comparison with the problem setting in \cite{LeBlancPhD}, 
the reader can refer to \cite{Dibaji,DibIshiiIFAC}. 

\begin{figure}[t]
  \centering
  \subfigure[(2,2)-robust]{%
      \label{fig:graph1}
      \includegraphics[width=32mm]{7node22robust}}
  \hspace*{4mm}
  \subfigure[Strongly connected]{%
      \label{fig:graph2}
      \includegraphics[width=32mm]{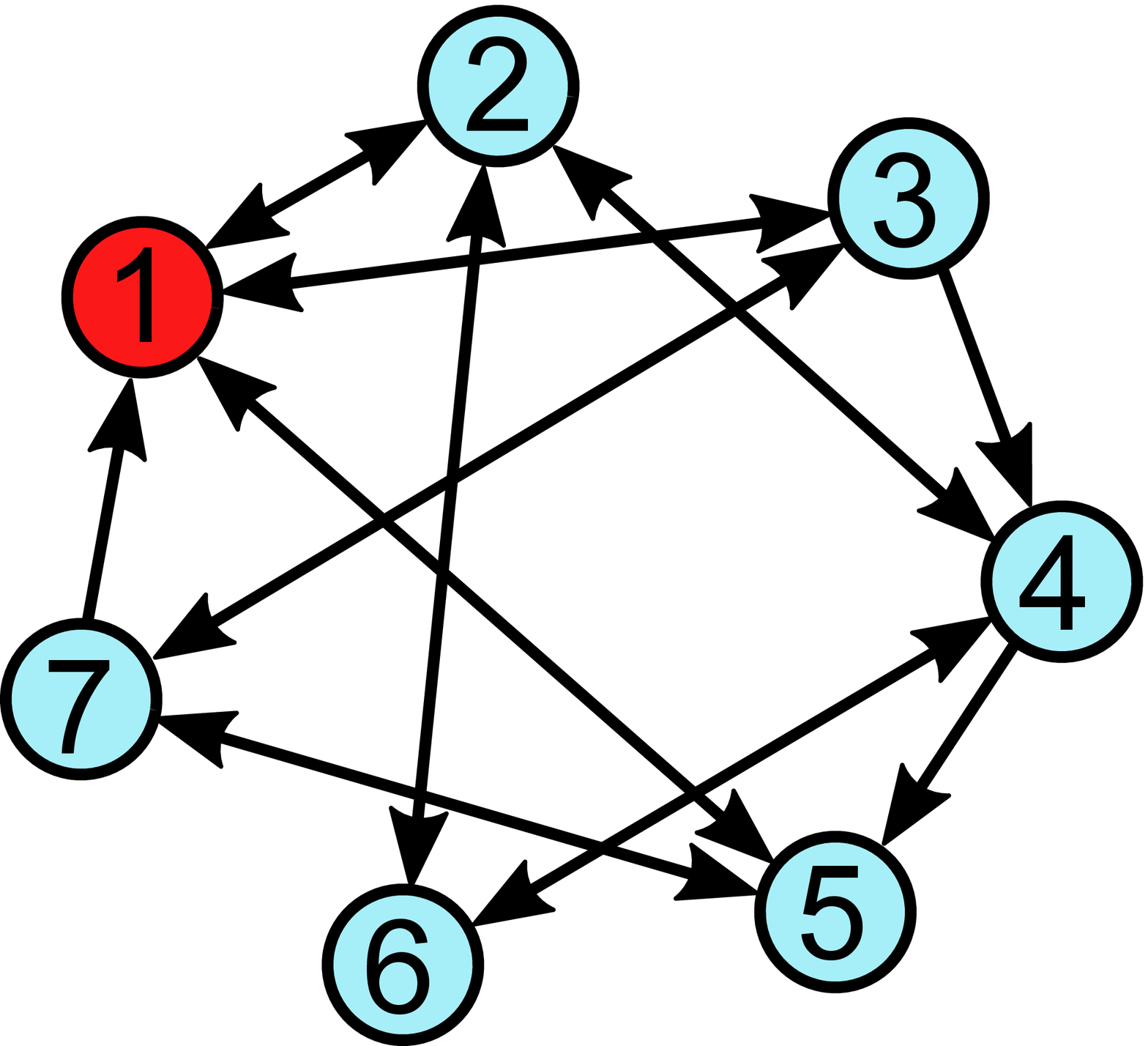}}\\
  \vspace*{6mm}
  \subfigure[3-robust]{%
      \label{fig:graph3}
      \includegraphics[width=32mm]{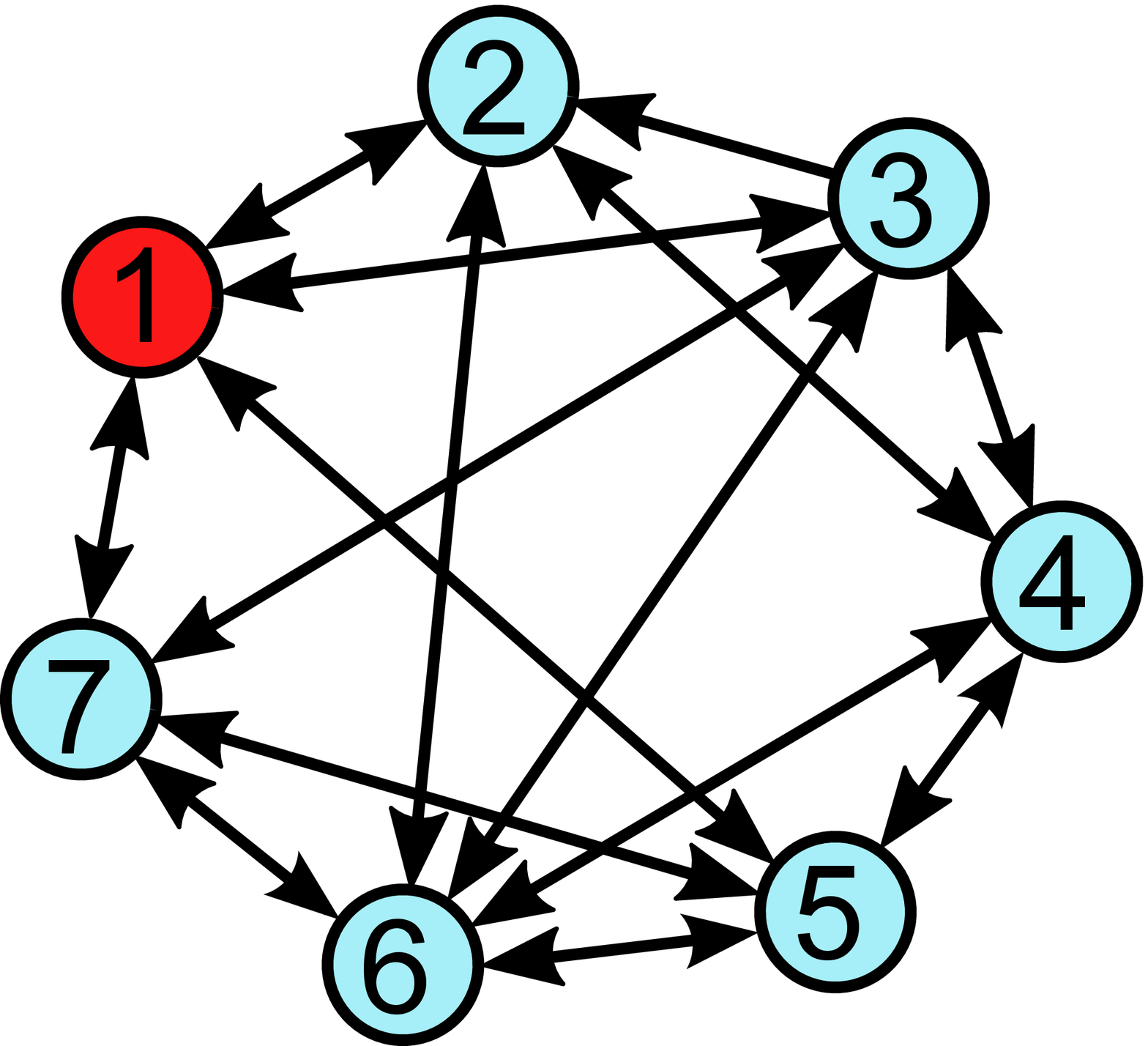}}
  \caption{Networks with seven agents}
\end{figure}

\section{Numerical Example}\label{Sect: Simulations}

In this section, we illustrate the effectiveness of the proposed resilient consensus algorithms through a numerical example. 

Consider the network with seven nodes in Fig.~\ref{fig:graph1}, 
which is the same as the one in Fig.~\ref{fig: 7node22robust}.
As discussed in Section~\ref{graphnot}, this graph is (2,2)-robust. 
Throughout the example, it is assumed that the agents start from 
the initial values $x[0]=[1~10~1~10~1~10~1]^T$. 
To misguide the normal agents, agent~$1$ is chosen to be malicious, 
and it alternates between two values 
as $x_1[2m]=1$ and $x_1[2m+1]=10$ for $m \in \mathbb{Z}_+$,
which are the values in $x^N[0]$.
\textcolor{black}{%
In the plots, the malicious agent's behavior is not shown as
it tends to move very frequently in the different scenarios.
The line colors used in the plots for the normal agents $2,3,\ldots,7$ 
are shown in Fig.~\ref{fig:color_code}.
}

\subsubsection{Synchronous Networks}
\begin{figure}[t]
	\centering
	\vspace*{0mm}
	\includegraphics[width=70mm]{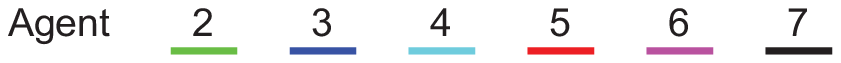}
	\vspace*{-2mm}
	\caption{\textcolor{black}{%
                 Line colors of the normal agents in the plots.}}
	\label{fig:color_code}
	\vspace*{4mm}
	\includegraphics[width=95mm,height=3.6cm]{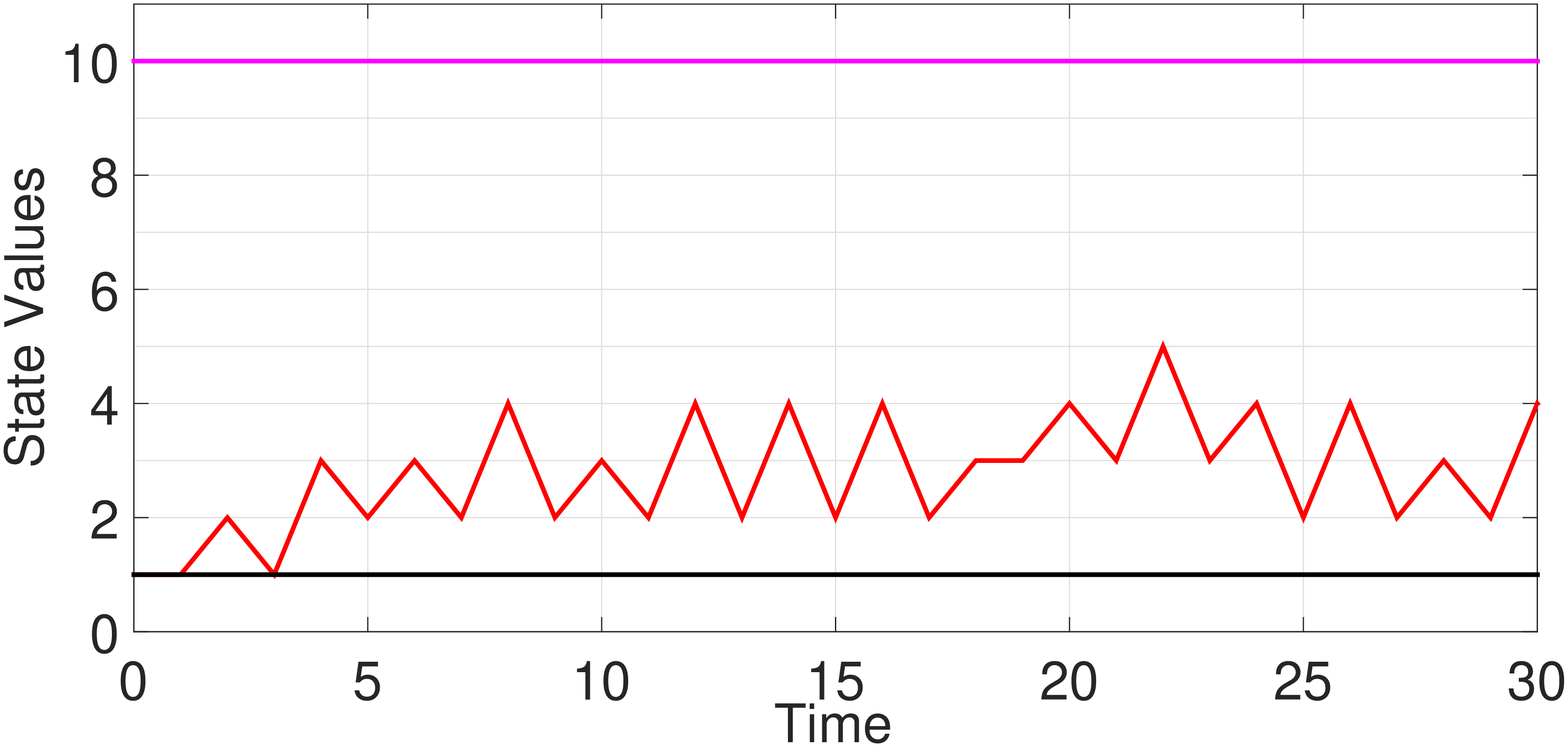}
	\vspace*{-6mm}
	\caption{Synchronous QW-MSR algorithm over a nonrobust graph.}
	\label{fig: Synchnotrobust}
	\vspace*{4mm}
	\includegraphics[width=95mm,height=3.6cm]{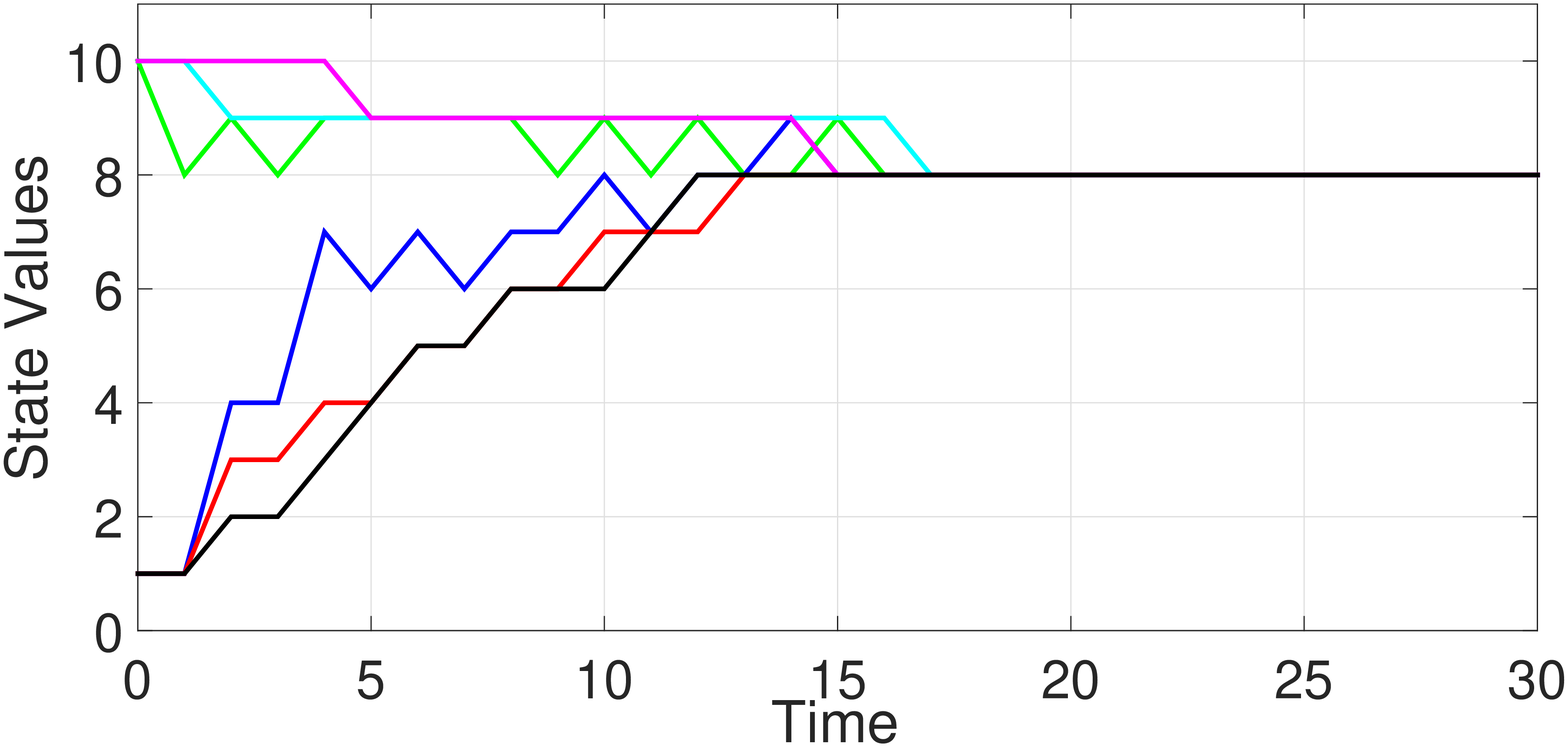}
	\vspace*{-6mm}
	\caption{Synchronous QW-MSR algorithm over a $(2,2)$-robust graph}
	\label{fig: Synchrobust}
\end{figure}

First, we conducted simulations on the graph in Fig.~\ref{fig:graph2}, 
which is constructed by removing 
the edges $(1,7)$, $(3,2)$, $(5,6)$, $(6,3)$, and $(6,7)$ from the original
graph $\mathcal{G}$ in Fig.~\ref{fig:graph1}. 
The obtained graph is no longer (2,2)-robust, but is strongly connected. After applying the 
QW-MSR algorithm on the network, the normal agents are deceived by agent 1 and do not reach agreement. Fig.~\ref{fig: Synchnotrobust} illustrates how the normal agents make updates under this configuration. As it is seen, while agents~2, 4, and 6 stay at~10 and agents~3 and~7 
stay at~1, agent~5 fluctuates between these two values. 
This shows that for resilient consensus, strong connectivity may not be sufficient. 
It is noteworthy that by removing the malicious node~1, the network of normal agents 
has a spanning tree. 
Hence, although the normal agents form a graph satisfying 
the well-known necessary and sufficient condition for reaching consensus, 
the adversarial effect of the malicious agent prevents agreement among them.

Next, we made simulations for the original $(2,2)$-robust graph $\mathcal{G}$ in 
Fig.~\ref{fig:graph1}. 
As shown in Fig.~\ref{fig: Synchrobust}, 
the normal agents after 17 steps meet at $x^*=8$.
This confirms the result of Theorem~\ref{Thm: NecSufSYNCHNONDELAY}
for the case of synchronous interactions.

\subsubsection{Asynchronous Networks without Delay}
\begin{figure}[t]
	\centering
	\vspace*{0mm}
	\includegraphics[width=95mm,height=3.6cm]{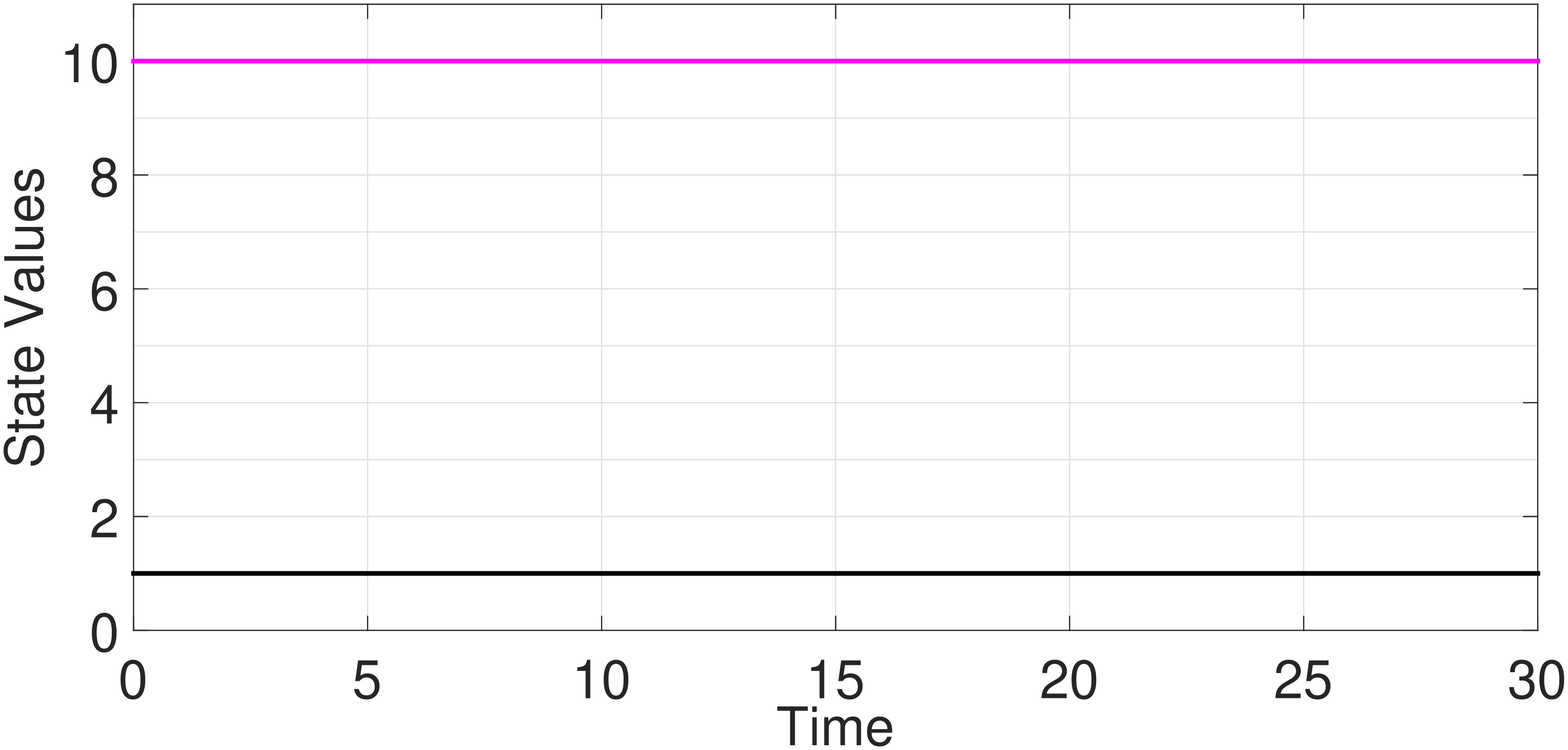}
	\vspace*{-5mm}
	\caption{Asynchronous QW-MSR algorithm over a $(2,2)$-robust graph 
             with deterministic update times.}
	\label{fig: Deterasyncnotrobust}
	\vspace*{4mm}
	\includegraphics[width=95mm,height=3.6cm]{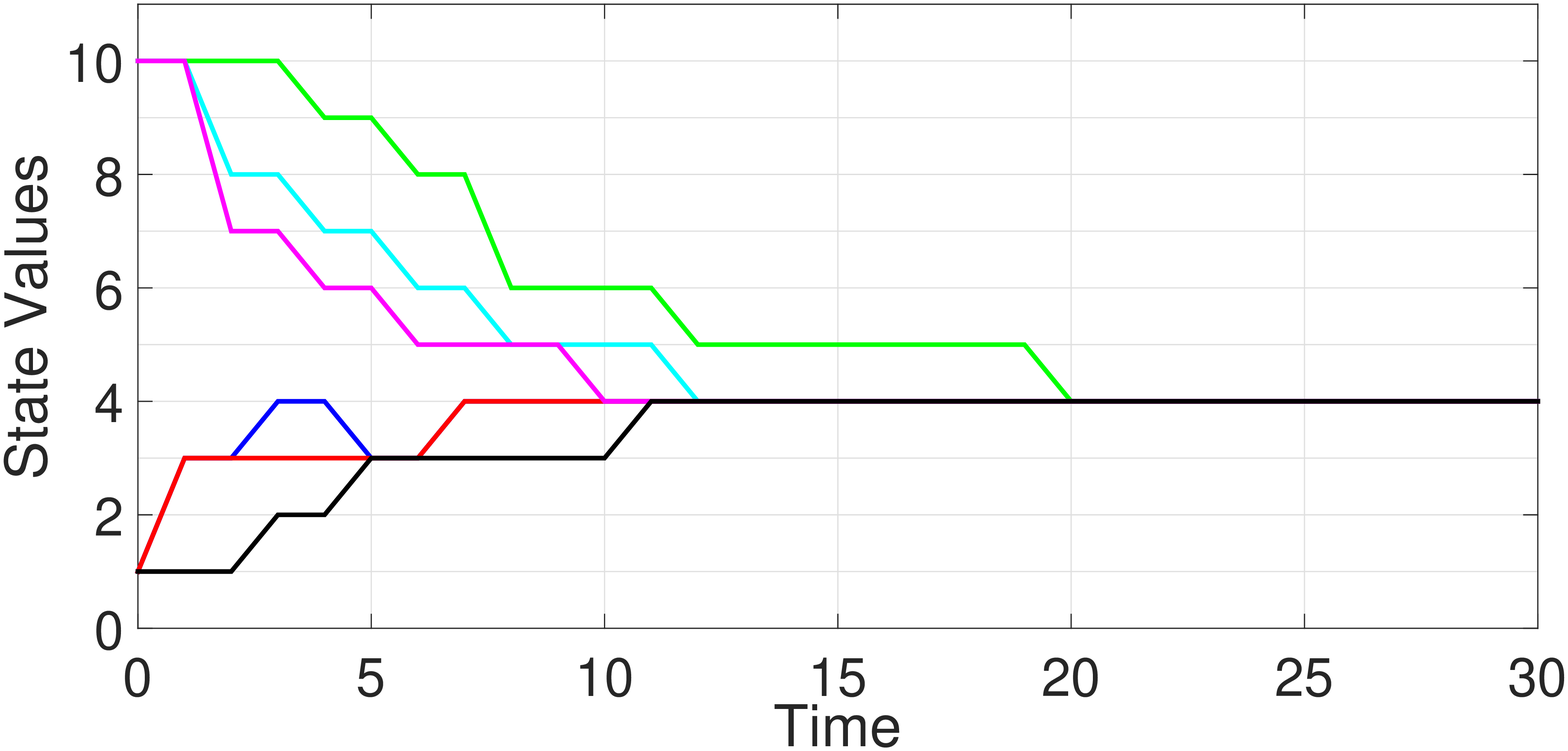}
	\vspace*{-5mm}
	\caption{Asynchronous QW-MSR algorithm over a $3$-robust graph 
             with deterministic update times.}
	\label{fig: Deterasyncrobust}
	\vspace*{4mm}
	\includegraphics[width=95mm,height=3.6cm]{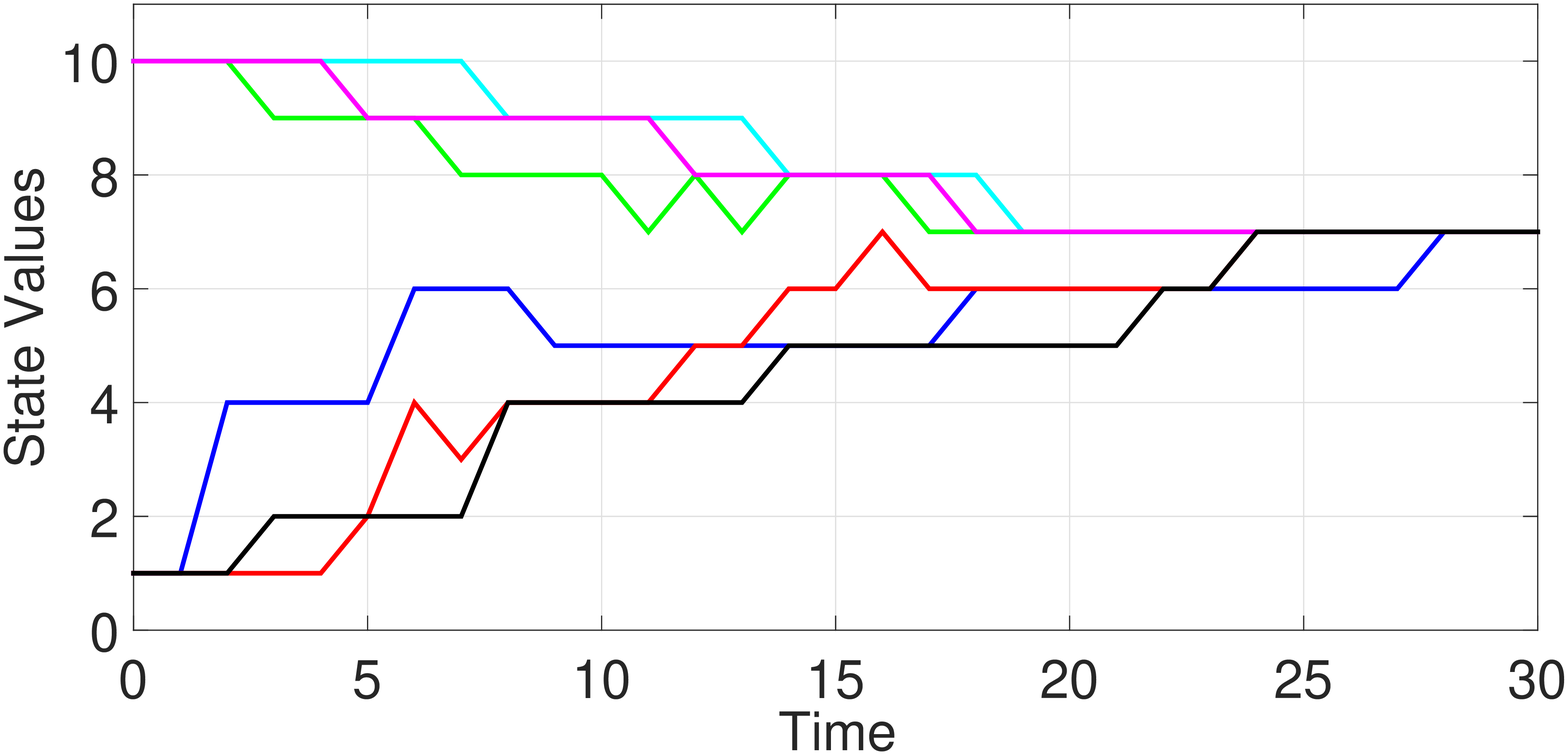}
	\vspace*{-5mm}
	\caption{Asynchronous QW-MSR algorithm over a $(2,2)$-robust graph 
             with probabilistic update times.}
	\label{fig: Randomasyncrobust}
\end{figure}

Then, we carried out simulations for the case where normal agents make asynchronous updates.
First, we examined a deterministic rule by assigning different updating times
for agents~3, 5, and~7, whose initial values are 1, and agents~2, 4, and~6, 
whose initial values are 10, as follows:
\begin{equation}
  \mathcal{U}[k]
    = \begin{cases}
         \{3,5,7\}&~\text{if $k=2m$},\\
         \{2,4,6\}&~\text{if $k=2m+1$}
      \end{cases}
\label{Eqn: UpdatesSimulation}
\end{equation}
for $m \in \mathbb{Z}_+$. 
After applying the QW-MSR algorithm on the original $\mathcal{G}$ 
in Fig.~\ref{fig:graph1}, the normal agents do not change their state values at all 
as shown in Fig.~\ref{fig: Deterasyncnotrobust}. Note that the underlying graph 
satisfies the necessary condition, but not the sufficient condition. 

In the next simulation, we examined the sufficient condition stated in 
Theorem~\ref{Thm: SufCondDeterminiticAsynchNONDELAY} by modifying $\mathcal{G}$ 
to become 3-robust. 
We added the edges $(3,6)$, $(4,3)$, $(5,4)$, $(6,5)$, and $(7,6)$ to 
$\mathcal{G}$ to this aim and obtained the graph in Fig.~\ref{fig:graph3}.
The result is seen in Fig.~\ref{fig: Deterasyncrobust} where the normal 
agents agree upon $x^*=4$ after 20 time steps. 

As the last step, we employed the probabilistic update rule over the 
$(2,2)$-robust graph in Fig.~\ref{fig:graph1}, where each normal agent 
makes updates with probability $p=0.5$. The result is shown in 
Fig.~\ref{fig: Randomasyncrobust}. We observe that the normal agents 
reach $x^*=7$ after 28 time steps. This verifies that randomization 
in the update times is capable to relax the required robustness of the network. 
However, compared to Fig.~\ref{fig: Deterasyncrobust}, it takes longer 
to reach its final value, which may be
because of the sparser network structure.



\subsubsection{Asynchronous and Delayed Networks with Deterministic Updates}
Here, we performed simulations considering delays in the original network 
$\mathcal{G}$ in Fig.~\ref{fig:graph1}. 
The updates follow the same deterministic rule in \eqref{Eqn: UpdatesSimulation}.
The communication delays are 
present in the edges from agent~$1$ to its neighbors and are set as below:
\begin{equation}\label{Eqn: DelaysSimulation}
  (\tau_{i1}[2m],\tau_{i1}[2m+1])
   = \begin{cases}
       (7,8)&~\text{if $i=2$},\\
       (8,7)&~\text{if $i=3,5,7$}
     \end{cases}
\end{equation}
for $m\in\mathbb{Z}_+$. 
Although the underlying graph meets the necessary condition stated 
in Proposition~\ref{Thm: NecCond}, 
none of the normal agents changes its value and their 
responses look the same as those in Fig.~\ref{fig: Deterasyncnotrobust}.
Thus, again, they form two clusters at $1$ and $10$ 
and fail to reach consensus.


Next, we examined Theorem~\ref{Thm: SufCondDeterDELAY} 
by using the $3$-robust graph in Fig.~\ref{fig:graph3}.
This graph was obtained by adding the edges 
$(3,6)$, $(4,3)$, $(5,4)$, $(6,5)$, and $(7,6)$ to $\mathcal{G}$.
In the network, 
the updating times \eqref{Eqn: UpdatesSimulation}
and the time delays \eqref{Eqn: DelaysSimulation}
remain the same as in the previous simulations. 
The time responses are given in Fig.~\ref{fig: Deterministicrobust}, 
where the normal agents agree upon $x^*=5$ after 19 time steps. 
Thus, we have verified Theorem~\ref{Thm: SufCondDeterDELAY}.

\subsubsection{Asynchronous and Delayed Networks with Randomized Updates}
\begin{figure}[t]
	\centering
	\vspace*{-.5mm}
	\includegraphics[width=90mm,height=3.75cm]{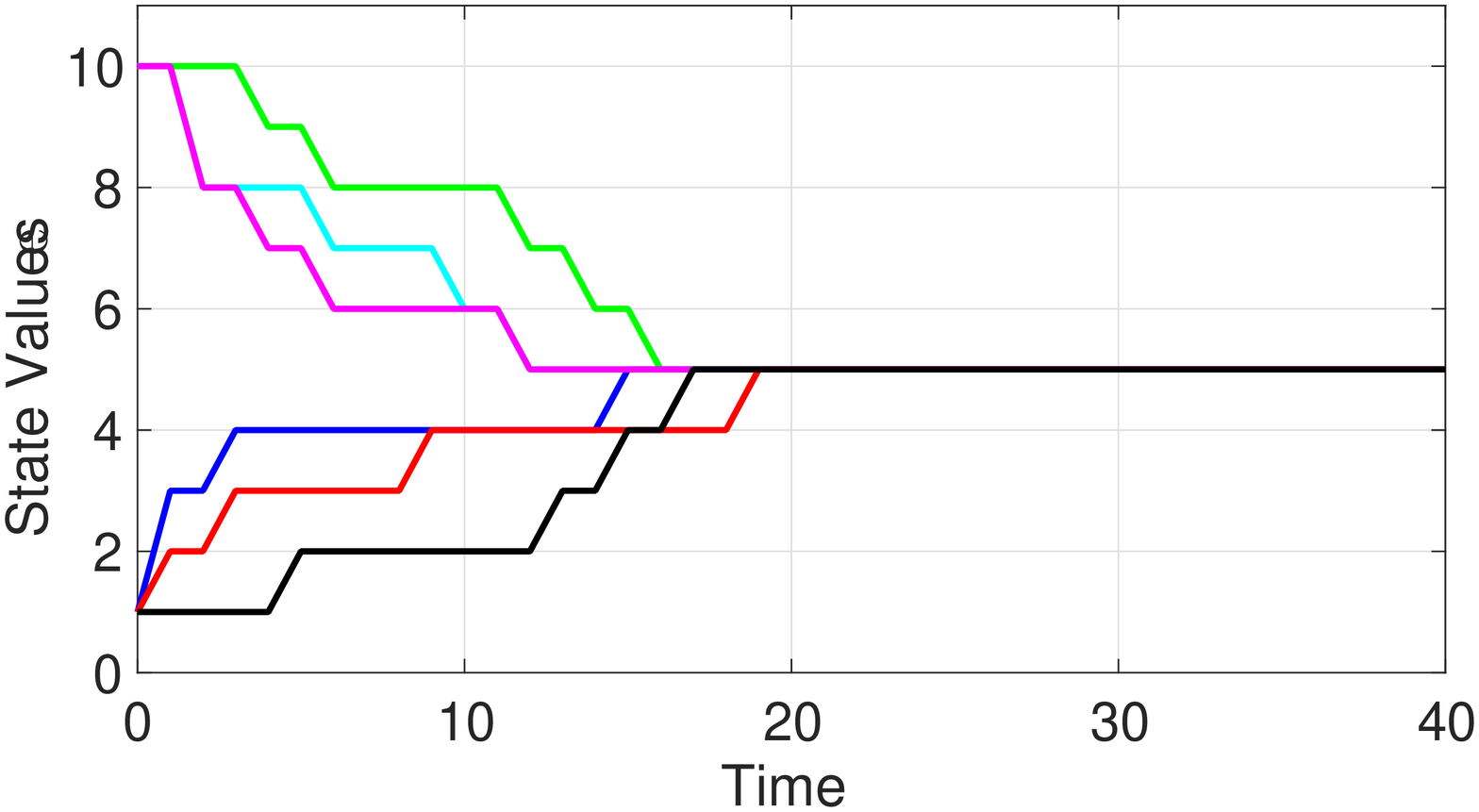}
	\vspace*{-5.8mm}
	\caption{Asynchronous QW-MSR algorithm over a $3$-robust graph under delays and deterministic updates ($f=1$).}
	\label{fig: Deterministicrobust}
	\vspace*{3.7mm}
	\includegraphics[width=90mm,height=3.72cm]{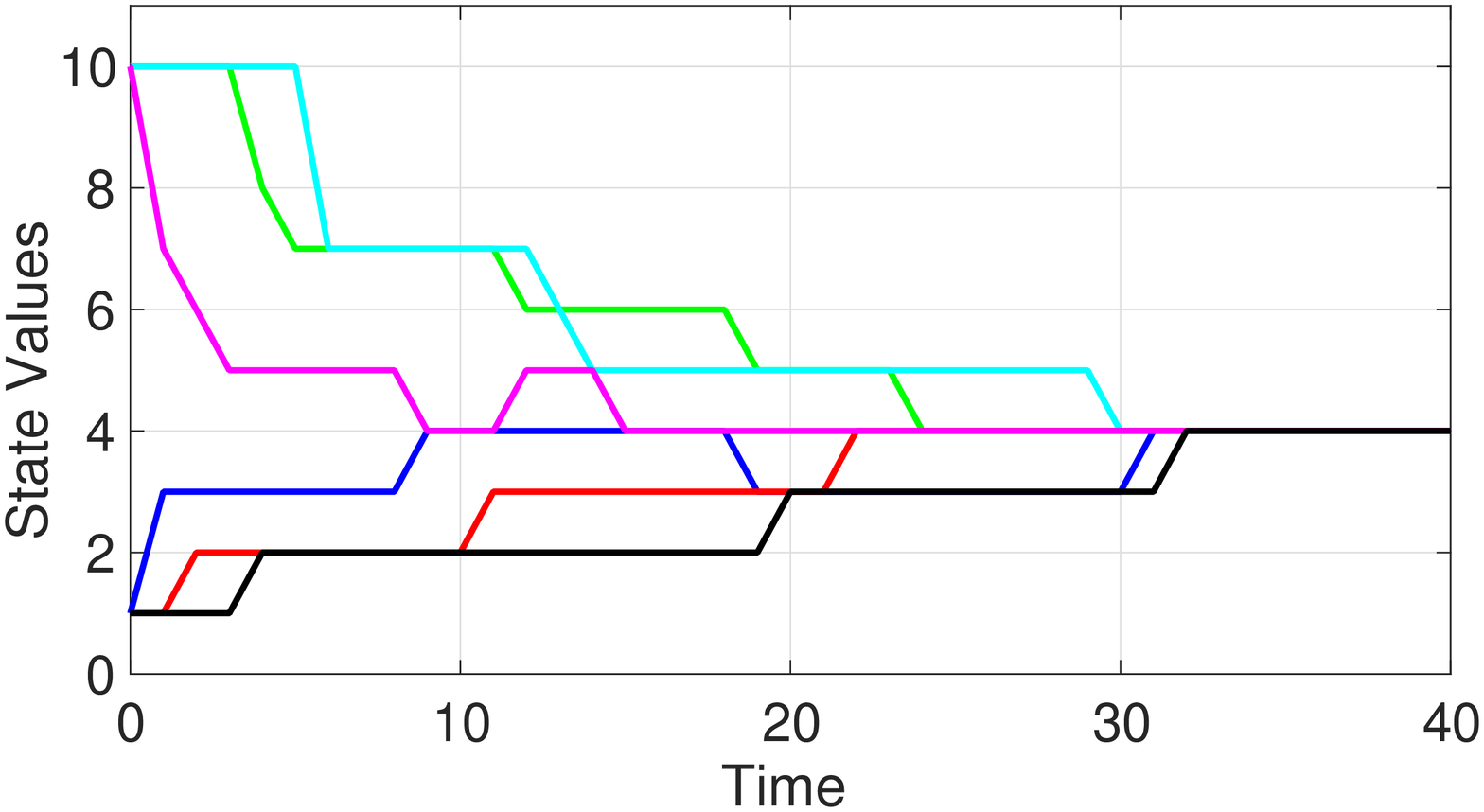}
	\vspace*{-5.8mm}
	\caption{Asynchronous QW-MSR algorithm over a $3$-robust graph under delays and randomized updates ($f=1$).}
	\label{fig: randomizedrobust}
\end{figure}

As the last simulation study,
we checked if the randomization is helpful in the delayed information case.
We set each normal agent $i$ to be updating (i.e., $i\in\mathcal{U}[k]$) 
with probability $p_i=0.4$. The delays are defined with the rule 
\eqref{Eqn: DelaysSimulation}. After applying the asynchronous QW-MSR algorithm on the $3$-robust graph constructed in the previous example, we observe in Fig.~\ref{fig: randomizedrobust} that the normal agents meet at $x^*=4$ but with a lower speed at $k=32$. This corresponds to the result of Proposition~\ref{Prop: NecSufCondRandomAsynchDELAY}.

Finally, we employed the randomized update rules over the original $\mathcal{G}$. We could see that the normal agents form two groups at $1$ and $10$ exactly as in Fig.~\ref{fig: Deterasyncnotrobust}, which again verifies that the sufficient condition in Theorem~\ref{Thm: SufCondDeterDELAY} cannot be relaxed even if the update times are randomized.

As a result of these examples, we see that the malicious 
agent tries to hide its jumps (from $1$ to $10$ and vice versa) 
by imposing specific delays in the links to the normal agents. 
By this strategy, it successfully deceives the normal agents to stay 
at their values. This is independent of 
update times and even randomization in update times 
is not helpful. Thus, the network must be more robust 
in order to reach consensus.


\section{Conclusion}\label{Sect: Conclusion}

In this paper, the problem of quantized consensus in the presence 
of malicious agents 
has been considered. We have studied several classes of resilient
distributed algorithms that perform under different conditions such 
as synchronous and asynchronous schemes with deterministic 
and probabilistic updating times, and delayed information.
Necessary conditions and sufficient conditions for reaching consensus 
among non-faulty agents have been derived based on the notion
of graph robustness. 

In particular, we have made use of randomization in quantization 
as well as in the update times, which turn out to be critical
in obtaining tight topological conditions on the underlying graph. 
On the other hand, in the asynchronous and/or delayed case, 
we have observed how malicious agents can take advantage of the
properties in communication to prevent normal agents from 
reaching consensus. 

In future research, we intend to study convergence rates 
of the resilient algorithms 
(see Remark~\ref{remark:conv_time} for discussions)
as well as the application of MSR-type algorithms to 
other multi-agent problems with malicious information.



\section*{Acknowledgement}
The authors would like to thank Paolo Frasca and Shreyas Sundaram 
for the helpful discussions.

\section*{\textcolor{black}{%
Appendix: Proof of Theorem~\ref{Thm: SufCondDeterDELAY}}}

We must show that by applying the asynchronous QW-MSR to an $f$-total network, 
the conditions (C1)--(C3) in Lemma~\ref{lemma: Basisoftheproof} hold. 
First, we prove (C1), which is the safety condition.
Define the minimum and maximum values of the normal 
agents at time $k$ and the previous $\overline{\tau}$ time steps by
\begin{align}
\nonumber 
  \underline{z}[k] 
   &= \min (x^N[k], x^N[k-1], \ldots , x^N[k-\overline{\tau}]),\\
  \overline{z}[k] 
   &= \max (x^N[k], x^N[k-1], \ldots , x^N[k-\overline{\tau}]).
\label{eqn:z_min_max}
\end{align}
In the network, there are at most $f$ malicious agents, and, at each time step, 
each normal agent removes the values of at most $f$ neighbors from the above and below. 
Hence, those faulty agents with unsafe values outside the
interval $\bigl[\underline{z}[k],\overline{z}[k]\bigr]$ are all ignored at each time step. 
In other words, each normal agent $i$ is affected only by values within
$\bigl[\underline{z}[k],\overline{z}[k]\bigr]$. 
It thus follows that the value of the normal agent $i$ 
in the update rule \eqref{eqn: ProbQuanGenUpdateDELAY} 
is upper bounded by
\begin{align}%
 x_i[k+1] 
  &
  \leq \Big\lceil 
           \sum_{j\in \mathcal{N}_i\cup\{i\}}
            w_{\tau,ij}[k] x_j[k-\tau_{ij}[k]]
       \Big\rceil\nonumber \\
  &\leq \Big\lceil 
          \sum_{j\in \mathcal{N}_i\cup\{i\}}  
            w_{\tau,ij}[k] \overline{z}[k]
       \Big\rceil 
  = \big\lceil 
      \overline{z}[k]
    \big\rceil 
  = \overline{z}[k] 
\label{eqn: UpperboundNormal}
\end{align}
with $\tau_{ii}=0$ for $i$.
Also, it is clear that
\begin{align*}
  &x_i[k] \leq \overline{z}[k],~~ x_i[k-1] \leq \overline{z}[k],\ldots,\\
  &x_i[k-(\overline{\tau}-1)]
    =x_i[k+1-\overline{\tau}] 
    \leq \overline{z}[k]  
\end{align*}
for $i\in\mathcal {V}\setminus\mathcal {M}$. Hence, with \eqref{eqn: UpperboundNormal}, we have 
\begin{align*}
 \overline{z}[k+1]
   &= \max\left(x^N[k+1],\cdots,x^N[k+1-\overline{\tau}]\right) 
   \leq \overline{z}[k].
\end{align*}
This implies that $\overline{z}[k+1] \leq \overline{z}[k]$, that is,
$\overline{z}[k]$ is a monotonically non-increasing function of time. 
Likewise, agent $i$'s value can be bounded from below as
\begin{align*}
  x_i[k+1] 
   &\geq \Big\lfloor 
            \sum_{j\in \mathcal{N}_i\cup\{i\}}  
              w_{\tau,ij}[k] x_j[k-\tau_{ij}[k]]
         \Big\rfloor \\
   &\geq \Big\lfloor 
            \sum_{j\in \mathcal{N}_i\cup\{i\}}  
              w_{\tau,ij}[k] \underline{z}[k]
         \Big\rfloor 
   = \big\lfloor 
       \underline{z}[k]
     \big\rfloor
   = \underline{z}[k].
\end{align*}
Thus, by similar arguments, we have 
$\underline{z}[k+1] \geq \underline{z}[k]$, which shows that
$\underline{z}[k]$ is a monotonically non-decreasing function of time. 
Consequently, we conclude that for the normal agent $i$, its state 
satisfies 
$x_i[k]\in\bigl[\underline{z}[k],\overline{z}[k]\bigr]\subset \mathcal{S}_{\overline{\tau}}$ 
for all $k$ with the interval $\mathcal{S}_{\overline{\tau}}$ 
given in \eqref{eqn: SafetyDELAY}.
Thus, (C1) is established. 

Next, we prove (C2) in Lemma~\ref{lemma: Basisoftheproof}. 
Since $\underline{z}[k]$ and $\overline{z}[k]$ are contained 
in $\mathcal{S}_{\overline{\tau}}$ and 
are monotone, there is a finite time $k_c$ such that they both reach their final values with probability 1. Denote the final values of $\underline{z}[k]$ and $\overline{z}[k]$ by $\underline{z}^*$ and $\overline{z}^*$, respectively. 
Now, to conduct a proof by contradiction, we assume $\underline{z}^* < \overline{z}^*$. Then, denote by $\mathcal{X}_{1\overline{\tau}}[k]$ the set of all normal agents 
at time $k \geq k_c$ with state values equal 
to $\overline{z}^*$. Likewise, denote by $\mathcal{X}_{2\overline{\tau}}[k]$ 
the set of normal agents that have the minimum state values 
$\underline{z}^*$. That is, 
\begin{equation}
\begin{split}
 \mathcal{X}_{1\overline{\tau}}[k] 
   &= \bigl\{i\in\mathcal{V}\setminus\mathcal{M}:~x_i[k]= \overline{z}^*\bigr\},\\
 \mathcal{X}_{2\overline{\tau}}[k] 
   &= \bigl\{i\in\mathcal{V}\setminus\mathcal{M}:~x_i[k]= \underline{z}^*\bigr\}.
\end{split}
\label{eqn:XDeterministic}
\end{equation}
From the convergence of $\overline{z}[k]$ to $\overline{z}^*$, 
we know that the sequence of $\mathcal{X}_{1\overline{\tau}}[k_c+\ell]$,
$\ell=0,\ldots,\overline{\tau}$, must collectively contain 
at least one normal agent that has the value $\overline{z}^*$, i.e.,
\begin{equation}
  \bigcup\limits_{k=k_c}^{k_c+\overline{\tau}}  
   \mathcal{X}_{1\overline{\tau}}[k] 
   \neq \emptyset
\label{eqn: NonemptinessofUnions}
\end{equation}
with probability 1. The same holds for $\underline{z}[k]$. 
We claim that in fact $\mathcal{X}_{1\overline{\tau}}[k_c]$ 
is nonempty. This is proven by showing that 
if $\mathcal{X}_{1\overline{\tau}}[k_c]$ is empty, 
then $\cup_{k=k_c}^{k_c+\overline{\tau}} 
\mathcal{X}_{1\overline{\tau}}[k] = \emptyset$ with 
non-zero probability, 
which is in contradiction with \eqref{eqn: NonemptinessofUnions}. 
To this end, it is enough to show that 
if $\mathcal{X}_{1\overline{\tau}}[k_c+\ell]$ is empty, 
then the probability of $\mathcal{X}_{1\overline{\tau}}[k_c+\ell+1]$ 
to be empty is non-zero. 

First, we show that none of the normal agents in 
$\mathcal{V} \setminus \mathcal{X}_{1\overline{\tau}}[k]$ enters 
$\mathcal{X}_{1\overline{\tau}}[k+1]$
at the next time step with positive probability. 
Assume that the normal agent $i$ makes an update at 
time $k_c+\ell$. Note that if there is not such an 
updating agent at time $k_c+\ell$, none of them can 
enter $\mathcal{X}_{1\overline{\tau}}[k_c+\ell+1]$. 
However, we know that each normal agent makes an 
update at least once in $\bar{\tau}$ time steps. We 
also know by the assumption on emptiness of 
$\mathcal{X}_{1\overline{\tau}}[k_c+\ell]$ that agent $i$ is 
upper bounded by $\overline{z}^*-1$. Thus,
\begin{equation}\label{Eqn: ProvingNonEmptiness}
   x_i[k_c+\ell+1] 
     \leq Q ((1-\beta)\overline{z}^*+\beta(\overline{z}^*-1))
     = Q (\overline{z}^*-\beta),
\end{equation}
where $Q (\overline{z}^*-\beta)=\overline{z}^*-1$ 
with probability $1-\beta$. The same arguments can be employed to 
prove that  $\mathcal{X}_{2\overline{\tau}}[k_c]$ is nonempty.

Then, we show that with positive probability, at the next time step, 
the agents in $\mathcal{X}_{1\overline{\tau}}[k]$ decrease their values, 
and the agents in $\mathcal{X}_{2\overline{\tau}}[k]$ increase 
their values. 
The sets $\mathcal{X}_{1\overline{\tau}}[k_c]$ and 
$\mathcal{X}_{2\overline{\tau}}[k_c]$ are disjoint and 
nonempty; therefore, we can make use of $(2f+1)$-robustness 
of the underlying graph $\mathcal{G}[k]$. At least one of 
these two sets includes a node with $(2f+1)$ incoming links 
from outside. We show this for the case the normal agent $i$ 
is in $\mathcal{X}_{1\overline{\tau}}[k]$; the other case with 
$\mathcal{X}_{2\overline{\tau}}[k]$ can be established similarly. 
The value of this agent clearly is $x_i[k]=\overline{z}^*$.

In step 2 of the asynchronous QW-MSR, agent $i$ makes an update using 
at least one agent from its neighbors whose values are smaller 
than $\overline{z}^*$. It also neglects all values larger 
than $\overline{z}^*$ since there are at most $f$ such values. 
Thus, by the update rule 
\eqref{eqn: ProbQuanGenUpdateDELAY}, we write 
\begin{equation}\label{eqn: ShrinkingSets2}
x_i[k+1] 
 \leq Q\bigl( 
          (1-\beta)\overline{z}^*+\beta(\overline{z}^*-1) )
  = Q ( \overline{z}^*-\beta ). 
\end{equation}
Here, by \eqref{eqn:Q},
the quantizer output takes the truncated value as 
$Q( \overline{z}^*-\beta)= \overline{z}^*-1$ with probability $1-\beta$. 
This indicates that with positive probability, one of the normal agents 
taking the maximum value $\overline{z}^*$
decreases its value by at least one. 
Similarly, if the normal agent $i$ is in $\mathcal{X}_{2\overline{\tau}}[k]$, 
then with positive probability, it chooses the ceil quantization, in which case
its value will increase above $\underline{z}^*$.

From the above discussion, we conclude that at each 
time step, some agents in $\mathcal{X}_{1\overline{\tau}}[k]$ 
or $\mathcal{X}_{2\overline{\tau}}[k]$ make updates 
and they decrease or increase, respectively, their values with 
positive probability. Hence, for any $k \geq k_c +\bar{k}\cdot n_N$, 
the number of normal agents in one of the sets 
$\mathcal{X}_{1\overline{\tau}}[k]$ and $\mathcal{X}_{2\overline{\tau}}[k]$ 
is zero with positive probability because there are 
only $n_N$ such agents and each makes updates at least 
once in $\bar{k}$ time steps. This is a contradiction 
and proves (C2).

Finally, we must show (C3) in Lemma~\ref{lemma: Basisoftheproof}.
For this step, we can follow along similar lines as 
this part in the proof of Theorem~\ref{Thm: NecSufSYNCHNONDELAY}.
After all normal agents reach consensus at the value $x^*$, 
when they make updates at time $k$, other agents $j$ taking values $x_j[k] \neq x^*$ are ignored. 
Thus, the third step of asynchronous QW-MSR
with the update rule in \eqref{eqn: ProbQuanGenUpdateDELAY}
results in $x_i[k+1]=x^*$ for $i \in \mathcal{V} \setminus \mathcal{M}$. 
Having shown (C3), we conclude the proof.
\hfill \mbox{$\blacksquare$}

\bibliographystyle{amsplain}

\end{document}